\documentclass[12pt]{article}
\usepackage{latexsym,amssymb,amsmath,slashed}
\newcommand{\ft}[2]{{\textstyle\frac{#1}{#2}}}

\def\Re{\mathop{\rm Re}\nolimits}
\def\Im{\mathop{\rm Im}\nolimits}

\def\rmi{{\rm i}}
\def\rmd{{\rm d}}

\newcommand{\hc}{{\rm h.c.}}
\newcommand{\bbox}{\lower.2ex\hbox{$\Box$}}
\newcommand{\SU}{\mathop{\rm SU}}

\newcommand{\U}{\mathop{\rm {}U}}

\csname @addtoreset\endcsname{equation}{section}
\usepackage{ifpdf}
\ifx\pdfoutput\undefined
   \pdffalse
   \usepackage{cite}
 \else
   \pdfoutput=1
   \pdftrue
  \usepackage[pdftex]{hyperref}
\fi
\usepackage{mciteplus}
\newcommand{\rf}[1]{(\ref{#1})}
\usepackage{color}

\def\be{\begin{equation}}
\def\ee{\end{equation}}
\newcommand{\ba}{\begin{eqnarray}}
\newcommand{\ea}{\end{eqnarray}}

\newcommand{\lp}{\left(}
\newcommand{\rp}{\right)}

\renewcommand{\O}{\Omega}
\newcommand{\C}{\mathcal{C}}
\newcommand{\Z}{\mathcal{Z}}
\renewcommand{\H}{\mathcal{H}}
\newcommand{\K}{\mathcal{K}}
\newcommand{\B}{\mathcal{B}_\mu}
\renewcommand{\L}{\Lambda}
\newcommand{\D}{\mathcal{D}}
\newcommand{\V}{\mathcal{V}}
\newcommand{\tC}{\tilde{\mathcal{C}}}
\newcommand{\tZ}{\tilde{\mathcal{Z}}}
\newcommand{\tH}{\tilde{\mathcal{H}}}
\newcommand{\tK}{\tilde{\mathcal{K}}}
\newcommand{\tB}{\tilde{\mathcal{B}}_\mu}
\newcommand{\tL}{\tilde{\Lambda}}
\newcommand{\tD}{\tilde{\mathcal{D}}}
\newcommand{\vp}{\varphi}
\newcommand{\Op}{\Omega^\Phi}
\def\rmi{{\rm i}}

\begin{document}

\begin{titlepage}
\begin{flushright}
CERN-TH-2016-052\\
TUW-16-05
\end{flushright}
\vspace{0.5cm}
\hskip 0.5cm
\begin{center}
\baselineskip=16pt

\newcommand{\Nn}{N}

{\LARGE {\bf Linear Versus Non-linear Supersymmetry,

\

in General}}

\

\

\

 { \large  \bf Sergio Ferrara$^{1,2,3}$},  {\large  \bf Renata Kallosh$^{4}$},  \\

\vskip 0.5 cm

   {\large \bf Antoine Van Proeyen$^5$}, {\large  \bf Timm Wrase$^6$} \vskip 0.8cm
{\small\sl\noindent
$^1$ Theoretical Physics Department, CERN CH�1211 Geneva 23, Switzerland\\\smallskip
$^2$ INFN - Laboratori Nazionali di Frascati Via Enrico Fermi 40, I-00044 Frascati, Italy\\\smallskip
$^3$ Department of Physics and Astronomy, U.C.L.A., Los Angeles CA 90095-1547, USA\\\smallskip
$^4$ SITP and Department of Physics, Stanford University, Stanford, California
94305 USA \\\smallskip
$^5$  KU Leuven, Institute for Theoretical Physics, Celestijnenlaan 200D, B-3001 Leuven,
Belgium \\\smallskip
$^6$ Institute for Theoretical Physics, TU Wien, Wiedner Hauptstr. 8-10, A-1040 Vienna, Austria}

\end{center}


\vskip 1cm
\begin{center}
{\bf Abstract}
\end{center}
{\small We study  superconformal and supergravity models with constrained superfields. The underlying version of such models with all unconstrained superfields and linearly realized supersymmetry is presented here, in addition to the physical multiplets there are Lagrange multiplier (LM) superfields. Once the equations of motion for the LM superfields are solved, some of the physical superfields become constrained.  The  linear supersymmetry of the original models becomes non-linearly realized, its exact form can be deduced from the original linear supersymmetry.  Known examples of constrained superfields are shown to require the following LM's: chiral superfields,  linear superfields, general complex superfields, some of them are multiplets with a spin.

}\vspace{2mm} \vfill \hrule width 3.cm \vspace{1mm}
{\footnotesize \noindent Sergio.Ferrara@cern.ch, kallosh@stanford.edu,
antoine.vanproeyen@fys.kuleuven.be, \\ timm.wrase@tuwien.ac.at }
\end{titlepage}
\addtocounter{page}{1}
 \tableofcontents{}
\section{Introduction}

Constrained superfields are useful in cosmology, both for the description of dark energy via de Sitter supergravity, as well as for inflationary model building. However, it is often believed that one can use non-linearly realized supersymmetry with constrained superfields only at low energies, otherwise one may encounter a violation of unitarity. A simple example of such a situation is when in supergravity in flat space the scalar component of a chiral superfield is very heavy. As shown in \cite{Casalbuoni:1988sx}, at energies above  $s_c=6 \sqrt{2\pi} \, m_{3/2} \, M_{Pl}$, unitarity may be violated. During inflation the relevant parameter is not the gravitino mass $m_{3/2}$ but a combination of it with the Hubble parameter, which sets the scale of spontaneous supersymmetry breaking $F \sim \sqrt{m_{3/2}^2 +H^2}\, M_{Pl}$, as shown in  \cite{Kallosh:2000ve}. Therefore, generically,   it was argued in \cite{Ferrara:2015tyn,Carrasco:2015iij}, that no violation of the unitarity bound is expected during inflation at sub-Planckian energy density. The issue of the unitarity bound during inflation was also raised in the context of inflationary models in \cite{Dall'Agata:2014oka} where it was  shown that during inflation no violation of unitarity takes place, due to specific features of these models.

However, during the exit from inflation and reheating as well as in particle physics, the situation with the unitarity bound still has to be investigated.  The experimental searches of supersymmetry are usually based on the assumption that super-partners have somewhat different masses.
 However,  if there are models with constrained superfields where the unitarity bound is not violated, maybe
 one can  speculate that in some cases the `linear superpartners' of  known particles are absent/extremely heavy?

The arguments above serve as a motivation for our study of  a consistency of models with constrained superfields where the action is invariant under a non-linearly realized local supersymmetry. This involves a derivation of these models from the underlying models with linear supersymmetry with additional multiplets, Lagrange multipliers, where all superfields are unconstrained.

The most famous example is the Volkov-Akulov (VA) model \cite{Volkov:1972jx, *Volkov:1973ix}, which has a spontaneously broken global supersymmetry and only a fermion field in the spectrum. The non-linear partner of the one fermion state is a two-fermions state, there are no bosons in the spectrum. The relation between linear and non-linear supersymmetries, often in the superspace context, was investigated  from the early days of supersymmetry, e.g. in  \cite{Rocek:1978nb, *Ivanov:1978mx, *Lindstrom:1979kq, *Samuel:1982uh, *Casalbuoni:1988xh}, where  the nilpotent chiral multiplet ${\bf S}^2=0$ where ${\bf S} \equiv S(x, \theta)$, was proposed to describe the VA theory.
Constrained superfields in global supersymmetry were suggested in the past and
many of them were described in \cite{Komargodski:2009rz}. The theory with a constrained nilpotent superfield ${\bf S}^2=0$ was shown to be equivalent to the VA model  \cite{Kuzenko:2010ef}.
A superfield expression of the Volkov-Akulov-Starobinsky supergravity and the explicit bosonic part of it was proposed in
\cite{Antoniadis:2014oya}.

During the last couple of years the strategy to find a complete local supergravity action with non-linearly realized supersymmetry was proposed.
The first step \cite{Ferrara:2014kva} was to introduce a set of Lagrange multipliers in the superconformal theory  where the chiral multiplet superconformal calculus was used,   and the corresponding $F$-term action was supplemented by a Lagrange multiplier (LM) term.   The superconformal action in such case was defined as follows
\begin{equation}
{\cal L} =  [N (X,\bar X)]_D + [\mathcal{W}(X)]_F +  [f_{AB} (X) \bar \lambda ^A P_L \lambda^B]_F+ \left[\sum _k \Lambda^k  A_k(X )\right] _F\,,
\label{chiral}
\end{equation}
where the LM's $\Lambda^k$ are chiral superfields. { \it  All supersymmetries are linearly realized as long as the chiral LM's are present in the action.} When the equations of motion for the $\Lambda^k $ are solved then they lead to constraints on the chiral superfields:
\be
A_k(X )=0\,.
\ee
A specific example\footnote{In a global supersymmetry model the LM for the square of the chiral multiplet was introduced and studied in \cite{Kuzenko:2011tj}.}
of one such LM in a superconformal theory is
\begin{equation}
{\cal L} =  [N (X,\bar X)]_D + [\mathcal{W}(X)]_F + [f_{AB} (X) \bar \lambda ^A P_L \lambda^B]_F  + \left[\Lambda  (X^1 )^2\right] _F\,.
\label{symbL1}
\end{equation}
This is a linear superconformal model that becomes a theory of one nilpotent chiral multiplet  $(X^1)^2=0$, interacting with other multiplets, when the equations of motion for the LM $\Lambda$ are solved. Other such models are associated with holomorphic constraints like $X^1 X^2=0$ or $X^1  W_\alpha=0$. In this last case the LM is a chiral multiplet $\Lambda^\alpha$ with spin. We will discuss various LM superfields in detail. The underlying linear models for chiral constraints are defined by \rf{chiral}. A class of superfield expressions for the supergravity models with constrained curvature superfield,  their dual with the nilpotent superfield, and their bosonic actions were presented in \cite{Dudas:2015eha}.

A complete supergravity action with fermions and non-linearly realized local supersymmetry of the VA type, generalizing the global case to the so called `pure de Sitter supergravity', was presented in \cite{Bergshoeff:2015tra, *Hasegawa:2015bza} for one nilpotent superfield interacting with supergravity, and for a general coupling with supergravity and other multiplets in \cite{Kallosh:2015tea, *Schillo:2015ssx}. Another interesting aspect of the relation between linear and non-linear supergravities with a nilpotent multiplet was revealed in \cite{Kallosh:2015pho}, where it was shown that one can derive the model with non-linear supersymmetry by taking a formal limit in which the mass of the sgoldstino goes to infinity, starting from a model with a linearly realized supersymmetry. Examples of linear supersymmetry models that do not support constrained multiplets were reviewed in \cite{Ghilencea:2015aph}.  For the consistency of the non-linearly realized supersymmetry models it is possible but not necessary to derive them from linear models. The method of Lagrange multipliers of a general nature, which will be developed in this paper, is a direct tool for constructing consistent models of non-linear supersymmetry and constrained superfields.

 New orthogonal nilpotent superfields, which were recently used in the context of supergravity inflation in \cite{Kahn:2015mla,Ferrara:2015tyn,Carrasco:2015iij,Dall'Agata:2015lek}, are not described by the chiral LM's. The so-called  `relaxed set of constraints' is given by the requirement that
the inflaton superfield ${\bf \Phi}$ and the stabilizer superfield ${\bf S}$ satisfy the following constraint
 \be
 {\bf S}^2=0\, ,  \qquad  {\cal  D}_{\dot \alpha} ({\bf S } {\bf B})=0\,,
\label{rel} \ee
 where
\be
{\bf B}=\frac{1}{2\rmi}({\bf \Phi -\bar \Phi})\,.
\ee
 The orthogonal nilpotent constraints are
 \be
 {\bf S}^2=0\, ,  \qquad   {\bf S } {\bf B}=0\,.
\label{ort} \ee
We will show that for the `relaxed set of constraints' \rf{rel} the LM is a linear superfield, whereas for the  orthogonal nilpotent constraints \rf{ort} we will need a complex general superfield, as was already proposed in  \cite{Ferrara:2015tyn}. This requires us
to build superconformal models with complex superfields as LM's. In principle, the corresponding constructions may
be obtained from \cite{Kugo:1982cu,VanProeyen:1983wk, Kugo:1983mv}. Here we will present them in the framework and in the notation of \cite{Freedman:2012zz} and with some modifications useful for the analysis of the non-linear supersymmetry.

Thus, our goal is to define the superconformal models of the type
\begin{eqnarray}
{\cal L} ^{linear}= [N (X,\bar X)]_D + [\mathcal{W}(X)]_F +  [f_{AB} (X) \bar \lambda ^A P_L \lambda^B]_F+ \left[\sum _k \Lambda^k  A_k(X )\right] _F\nonumber\\ +\left[\sum _\ell \V^\ell  B_\ell(X, \bar X, \lambda^A,\cdots ) \right] _D\,.
\label{Dterm}
\end{eqnarray}
Here $\V^\ell$ are some linear or general complex superfields, LM's. {\it All supersymmetries are linearly realized as long as the LM's are present in the action \rf{Dterm}}. When the equations of motion for the $\V^\ell$ are solved, then the physical superfields satisfy the constraints
 \be
 B_\ell(X, \bar X, \lambda^A,\cdots )=0\,.
 \ee
Particular examples of such theories include the case of the `relaxed set of constraints' \rf{rel} and the orthogonal nilpotent  supermultiplets in \rf{ort},
studied in  the supergravity context in \cite{Ferrara:2015tyn,Dall'Agata:2015lek}.  An alternative method, not using LM's,  of relating linear supersymmetry and supergravity models to non-linear ones, is developed in \cite{Kallosh:2016hcm} for the constraints in  \rf{rel} and \rf{ort} by sending the masses of the sgoldstino,  inflatino and sinflaton  to infinity, as it was done before for the case of one nilpotent multiplet in \cite{Kallosh:2015pho}.

\section{Various constraints and Lagrange multipliers}

In this section, we will consider various constraints on multiplets. We consider here only the Lagrangian term that provides the constraint, hence the terms with $\Lambda ^k$ or $\V ^\ell$ in (\ref{Dterm}). We will consider the form of this term for various cases, and give the consistent Weyl weights for the constrained multiplets and the Lagrange multiplier multiplets. Note that there may be arguments for a particular Weyl weight from the way in which the fields appear in other parts of the action, and they can be modified by redefinitions of the form $X'= (S^0)^p X$, for $S^0$ the compensating chiral multiplet and $p$ some convenient power. Such a replacement just amounts to invertible field redefinitions.
We consider below chiral multiplets $X$, $Y$ and a gauge multiplet $W_\alpha $. The fields and possible (Weyl, chiral) weights are denoted as follows
\begin{equation}
  \begin{array}{ccc}
     \mbox{multiplet} & \mbox{fields} & \mbox{weights }(w,c)\mbox{ of defining field} \\ \hline
     X & \{X,\, P_L\Omega _X, \, F_X\} & (w,w)  \\
     Y & \{Y,\, P_L\Omega _Y, \, F_Y\} & (w',w')  \\
     W_\alpha  & \{P_L\lambda,\, V_\mu ,\,D\}  & (\ft32,\ft32)
   \end{array}
 \label{fieldsInConstraints}
\end{equation}
In each case, there is the constraint $X^2=0$. We demand that the rest of the Lagrangian is such that the auxiliary field
\begin{equation}
  F_X\neq 0\,.
 \label{FXnonzero}
\end{equation}
Other auxiliary fields can be zero or non-zero. In the latter case the Goldstino is a combination of various fermions. But in the following, we will take $\Omega _X$ to be the Goldstino.

We consider 5 models. The first 4 were already considered for rigid supersymmetry in  \cite{Komargodski:2009rz}. We give below the references to where the models were studied in supergravity.
\begin{equation*}
  \begin{array}{ccccccc}
  \mbox{model} & \mbox{constraint} & w,\,w' & \mbox{Dependent}& \mbox{Independent} & \mbox{Physical}& \mbox{References} \\ \hline
       & X^2=0 &  & X & \Omega _X,\,F_X & SUGRA & \\ \hline
    1 & XY=0 & &Y & \Omega _Y,\,F_Y & \Omega _Y & \cite{Dall'Agata:2015zla,Dall'Agata:2015lek} \\
    2 & X(Y- Y^*)=0 & w'=0 & \Im Y,\,\Omega _Y,\, F_Y& \Re Y & \Re Y &  \cite{Ferrara:2015tyn,Dall'Agata:2015lek} \\
     3 & X\,Y^* \mbox{ chiral} &w'=0 &\Omega _Y,\, F_Y & Y & Y & \cite{Ferrara:2015tyn,Dall'Agata:2015lek}\\
     3' & X\overline{{\cal D}}_{\dot \alpha }Y^*=0 & && &  & \\
     4 & XW_\alpha =0 & &\lambda & V_\mu ,\,D & V_\mu & \cite{Dall'Agata:2015zla,Ferrara:2015tyn,Dall'Agata:2015lek}   \\
     5 & X\, X^*{\cal D}_\alpha Y=0 & &\Omega _Y&Y,\,F_Y & Y& \cite{Dall'Agata:2015lek}
   \end{array}
\end{equation*}
Here, the column $w,\,w'$ gives restrictions on these weights in (\ref{fieldsInConstraints}), if there are any. `Dependent' fields are the fields that become functionals of other fields due to the constraint. The remaining fields are independent. After the super-Brout-Englert-Higgs mechanism and solving the field equations, the remaining fields are in the column denoted by `Physical'. Note that the first row is always included. This first row contains at the end the `pure de Sitter supergravity' model \cite{Bergshoeff:2015tra, *Hasegawa:2015bza}. The line 3' is an alternative formulation of the constraint 3. Model 2 is the one where only a real inflaton remains together with supergravity.

These constraints are implemented by the following Lagrange multiplier multiplets:
\begin{equation}
  \begin{array}{cccc}
  \mbox{model} & \mbox{LM} & \mbox{type} & ( w,\,c) \\ \hline
       & \Lambda _0 & \mbox{chiral} & (3-2w,\,3-2w) \\ \hline
     1 & \Lambda _1 & \mbox{chiral} & (3-w-w',\, 3-w-w') \\
     2 & \V _2 & \mbox{complex} & (2-w,\, -w) \\
     3 & \V _3 & \mbox{complex linear} & (2-w,\,-w) \\
     3' & \V_3 ^{\dot \alpha}  & \mbox{spinor} & (\ft32-w,\,-\ft32-w) \\
     4 & \Lambda _4^\alpha  & \mbox{chiral spinor} & (\ft32-w,\,\ft32-w) \\
     5 & \V _5^\alpha & \mbox{complex}& (\ft52-2w-w',\,\ft32-w')
   \end{array}
 \label{LMweights}
\end{equation}
3' is an alternative formulation of the linear multiplet, corresponding to $\V _3 = \overline{{\cal D}}_{\dot \alpha }\V_3 ^{\dot \alpha}$.

\section{General principles in short}
A complex multiplet corresponds to a complex superfield. In order that it can be upgraded to supergravity, we need consistency with the Weyl and chiral weights, such that it can be written in superconformal calculus. Complex multiplets have a first (complex) component ${\cal C}$, which can have (Weyl, chiral) weight $(w,c)$ arbitrary.\footnote{Fields in the superconformal algebra are defined by the eigenvalues of 4 Casimirs of $\SU(2,2)\times \U(1)$: $(w,c,J_1,J_2)$ (where $J_1$ and $J_2$ determine the Lorentz transformations of the fields). Limits for these weights are obtained in \cite{Ferrara:1999ed}. See also Appendix \ref{app:GeneralMult}.} Three noteworthy subcases are
\begin{eqnarray}
 \mbox{real multiplet} & : & {\cal C}={\cal C}^*\,,\qquad \mbox{if}\ c=0\,, \nonumber\\
 \mbox{chiral multiplet} & : & P_R{\cal Z}=0 \,,\qquad \mbox{if}\ c=w\,, \nonumber\\
 \mbox{antichiral multiplet} & : & P_L{\cal Z}=0 \,,\qquad \mbox{if}\ c=-w\,,
 \label{3examples}
\end{eqnarray}
where ${\cal Z}$ is the (second) fermionic component of the complex multiplet. We will give the detailed formulae below, including how to make general functions (e.g. products) of complex multiplets to construct other such multiplets. One just has to take functions that respect the homogeneity of the weights.

There are `density formulas' producing real actions, that are denoted as $[X]_F$, for $X$ the lowest component of a chiral multiplet, and $[{\cal C}]_D$ for ${\cal C}$ the lowest component of a
complex multiplet. These formulas require
\begin{eqnarray}
 [X]_F  & : & X \mbox{ has } (w,c)=(3,3)\,, \nonumber\\{}
 [{\cal C}]_D & : & {\cal C}\mbox{ has } (w,c)=(2,0) \,.
 \label{actionformulae}
\end{eqnarray}

Finally, an operation that is often used is to project a chiral multiplet out of a complex multiplet, which is the analogue of $\bar D^2$ in superspace. It is denoted as
\begin{equation}
  T({\cal C})\qquad \mbox{ for }{\cal C}\mbox{ having } (w,c)=(w,w-2)\mbox{ and then }T({\cal C})\mbox{ has weights} (w+1,w+1) \,.
 \label{Tdef}
\end{equation}

Many of these results were already given in \cite{Kugo:1982cu,VanProeyen:1983wk}.
Other operations are possible involving spinor multiplets, which are also treated in \cite{Kugo:1983mv}. We recapitulate the requirements on weights for these more general multiplets in Appendix \ref{app:GeneralMult}.

\section{Components and transformation rules of the complex multiplet}

The general complex scalar multiplet $\V$ has components\footnote{We redefined ${\cal H}$ and ${\cal K}$ w.r.t. \cite{Kugo:1982cu,VanProeyen:1983wk,Kugo:1983mv}. Our ${\cal H}$ is
in their terminology ${\cal H}+\rmi{\cal K}$ and our ${\cal K}$ is their ${\cal H}
-\rmi{\cal K}$. In this way our fields have definite chiral weights. Note that we use $\Lambda$ to denote the second to last component of the complex multiplet $\V$. It should be clear from the context whether we are referring to a chiral LM or the component of $\V$.}
\begin{equation}
  \left\{{\cal C},\,{\cal Z},\,{\cal H},\,{\cal K},\,{\cal B}_a ,\,\Lambda ,\,{\cal D}\right\}\,,
 \label{componentsComplexm}
\end{equation}
where ${\cal C}$, ${\cal H}$, ${\cal K}$ and ${\cal D}$ are complex scalars, and ${\cal Z}$ and $\Lambda $ are Dirac fermions.
The $Q$ and $S$-supersymmetry transformation laws are \cite{Kugo:1982cu}
\begin{eqnarray}
 \delta _{\epsilon,\eta } {\cal C} & = & \ft12\rmi \bar \epsilon \gamma _*{\cal Z} \,, \nonumber\\
  \delta _{\epsilon,\eta } P_L{\cal Z} & = & \ft12P_L\left( \rmi{\cal H}-\slashed{\cal B} -\rmi\slashed{\cal D}{\cal C}\right) \epsilon -\rmi (w+c)P_L\eta \,{\cal C}\,,\nonumber\\
 \delta _{\epsilon,\eta } P_R{\cal Z} & = & \ft12P_R\left( -\rmi{\cal K}-\slashed{\cal B} +\rmi\slashed{\cal D}{\cal C}\right) \epsilon +\rmi (w-c)P_R\eta \,{\cal C}\,,\nonumber\\
\delta_{\epsilon,\eta } {\cal H}&=&-\rmi\bar \epsilon P_R \left(\slashed{\cal D}{\cal Z}+\Lambda\right)+\rmi(w+c-2)\bar \eta P_L{\cal Z} \,,\nonumber\\
\delta_{\epsilon,\eta } {\cal K}&=&\phantom{-} \rmi\bar \epsilon P_L \left(\slashed{\cal D}{\cal Z}+\Lambda\right) +\rmi(-w+c+2) \bar \eta P_R{\cal Z} \,,\nonumber\\
\delta_{\epsilon,\eta } {\cal B}_a &=&-\ft12\bar \epsilon \left({\cal D}_a {\cal Z}+\gamma _a \Lambda \right)+\ft12\bar \eta (w+1+c\gamma _*)\gamma _a    {\cal Z}\,,\nonumber\\
\delta_{\epsilon,\eta } P_L\Lambda &=&\ft12\left[\gamma ^{ab }\left({\cal D}_a {\cal B}_b -\rmi{\cal D}_a {\cal D}_b {\cal C}\right)+\rmi{\cal D}\right]P_L\epsilon
-\ft12P_L\left(\rmi{\cal K}+\slashed{\cal B}+\rmi\slashed{\cal D}{\cal C}\right)(w+c\gamma _*)\eta \,,\nonumber\\
\delta_{\epsilon,\eta } P_R\Lambda &=&\ft12\left[\gamma ^{ab }\left({\cal D}_a {\cal B}_b +\rmi{\cal D}_a {\cal D}_b {\cal C}\right)-\rmi{\cal D}\right]P_R\epsilon
+\ft12P_R\left(\rmi {\cal H}-\slashed{\cal B}+\rmi\slashed{\cal D}{\cal C}\right)(w+c\gamma _*)\eta \,,\nonumber\\
\delta_{\epsilon,\eta } {\cal D}&=&\ft12\rmi\bar \epsilon \gamma _*\slashed{\cal D}\Lambda+\rmi\bar \eta (c+w\gamma _*)\left(\Lambda +\ft12 \slashed{\cal D}{\cal Z}\right) \,.
 \label{complexmult}
\end{eqnarray}
The (Weyl,chiral) weights of the fields are (complex conjugation changes the sign of $c$)
\begin{eqnarray}
 {\cal C} & : & (w,c)\,, \nonumber\\
 P_L{\cal Z} & : & (w+\ft12,c-\ft32)\,,\qquad P_R{\cal Z}\ :\ (w+\ft12,c+\ft32)\,,\nonumber\\
{\cal H}&:& (w+1,c-3)\,,\qquad{\cal K}\ :\ (w+1,c+3)\,,\qquad {\cal B}_a \ :\ (w+1,c)\,,\nonumber\\
P_L\Lambda  &:& (w+\ft32,c+\ft32)\,,\qquad P_R\Lambda \ :\ (w+\ft32,c-\ft32)\,,\nonumber\\
{\cal D}&:&(w+2,c)\,.
 \label{wcComplexMult}
\end{eqnarray}
They determine dilatations and chiral transformations for any field $\Phi $ according to
\begin{equation}
  \delta \Phi = w\,\lambda _{\rm D}\Phi + \rmi\, c \lambda _T\Phi \,.
 \label{delwcPhi}
\end{equation}
Finally, there are special conformal transformations for the following fields
\begin{eqnarray}
 \delta _{\rm K} {\cal B}_a  & = & -2\rmi c\,{\cal C}\lambda _{{\rm K}a }\,, \nonumber\\
 \delta _{\rm K}\Lambda  & = & (w+c\gamma _*)\slashed{\lambda }_{\rm K}{\cal Z}\,,\nonumber\\
 \delta _{\rm K} {\cal D}&=&2\left(w{\cal D}_a{\cal C}+2\rmi c \,{\cal B}_a \right)\lambda _{{\rm K} }^a \,.
 \label{Ktransfocomplex}
\end{eqnarray}

The covariant derivatives that appear in (\ref{complexmult}) are therefore
\begin{eqnarray}
 {\cal D}_\mu {\cal C} & = & \left(\partial _\mu -wb_\mu -\rmi c A_\mu \right)C-\ft12\rmi \bar \psi _\mu \gamma _*{\cal Z}\,, \nonumber\\
P_L {\cal D}_\mu {\cal Z} & = &  \left(\partial _\mu -(w+\ft12) b_\mu -\rmi (c-\ft32) A_\mu+\ft14\omega _\mu {}^{ab}\gamma _{ab} \right){\cal Z}\nonumber\\
&&-\ft12P_L\left( \rmi{\cal H}-\slashed{\cal B} -\rmi\slashed{\cal D}{\cal C}\right)\psi _\mu -\rmi (w+c)P_L \phi _\mu  {\cal C}\,,\nonumber\\
P_R {\cal D}_\mu {\cal Z} & = &  \left(\partial _\mu -(w+\ft12) b_\mu -\rmi (c+\ft32) A_\mu+\ft14\omega _\mu {}^{ab}\gamma _{ab} \right){\cal Z}\nonumber\\
&&-\ft12P_R\left( -\rmi{\cal K}-\slashed{\cal B} +\rmi\slashed{\cal D}{\cal C}\right)\psi _\mu  -\rmi (w-c)P_R\phi _\mu  {\cal C}\,,\nonumber\\
{\cal D}_a{\cal B}_b&=&e_a^\mu \left[\left(\partial _\mu -(w+1)b_\mu -\rmi cA_\mu \right)B_b+\omega _{\mu bc}B^c\right.\nonumber\\
&&\left.+\ft12\bar \psi _\mu \left({\cal D}_b{\cal Z}+\gamma _b\Lambda \right)
-\ft12\bar \phi _\mu (w+1+c\gamma _*)\gamma _b{\cal Z}+2\rmi c\,{\cal C}f_{\mu b}\right]\,,\nonumber\\
P_L{\cal D}_\mu \Lambda &=&P_L\left(\partial _\mu -(w+\ft32) b_\mu -\rmi (c+\ft32) A_\mu+\ft14\omega _\mu {}^{ab}\gamma _{ab} \right)\Lambda \nonumber\\
&&-\ft12\left[\gamma ^{ab }\left({\cal D}_a {\cal B}_b -\rmi{\cal D}_a {\cal D}_b {\cal C}\right)+\rmi{\cal D}\right]P_L\psi _\mu
+\ft12P_L\left(\rmi{\cal K}+\slashed{\cal B}+\rmi\slashed{\cal D}{\cal C}\right)(w+c\gamma _*)\phi _\mu\nonumber\\ &&-(w+c)\gamma _aP_R{\cal Z}f_\mu ^a\,,\nonumber\\
P_R{\cal D}_\mu \Lambda &=&P_R\left(\partial _\mu -(w+\ft32) b_\mu -\rmi (c-\ft32) A_\mu+\ft14\omega _\mu {}^{ab}\gamma _{ab} \right)\Lambda \nonumber\\
&&
 -\ft12\left[\gamma ^{ab }\left({\cal D}_a {\cal B}_b +\rmi{\cal D}_a {\cal D}_b {\cal C}\right)-\rmi{\cal D}\right]P_R\psi _\mu
-\ft12P_R\left(\rmi {\cal H}-\slashed{\cal B}+\rmi\slashed{\cal D}{\cal C}\right)(w+c\gamma _*)\phi _\mu \nonumber\\
&&-(w-c)\gamma _aP_L{\cal Z}f_\mu ^a  \,.
 \label{covdercomplex}
\end{eqnarray}
Furthermore, (\ref{complexmult}) contains the anticommutator of covariant derivatives on ${\cal C}$:
\begin{equation}
  {\cal D}_{[a}{\cal D}_{b]}{\cal C}=-\ft12\left[w\, R_{ab}({\rm D}) +\rmi c\,R_{ab}(T)\right]{\cal C}-\ft14\rmi\overline{{R}_{ab}(Q)}\gamma _*{\cal Z}\,.
 \label{DDC}
\end{equation}
The superconformal curvatures for the dilatation ${\rm D}$, $U(1)$ symmetry $T$, and supersymmetry $Q$, are defined in \cite{Freedman:2012zz}.

The multiplet that starts with the complex conjugate field ${\cal C}^*$ is
\begin{equation}
  \left\{{\cal C}^*,{\cal Z}^C,{\cal K}^*,{\cal H}^*,{\cal B}^*_\mu ,\Lambda^C ,{\cal D}^*\right\}\,,
 \label{componentscc}
\end{equation}
where $C$ denotes charge conjugation, which in a Majorana spinor representation is the complex conjugate.

\subsection{The restrictions to real and (anti)chiral}

The complex multiplet reduces to a \emph{real multiplet} when $C={\cal C}$ is real. That implies that its chiral weight vanishes $c=0$. Then ${\cal Z}= \zeta$ and $\Lambda $ are Majorana,
i.e. $(P_R{\cal Z})^C=P_L{\cal Z}$. Furthermore, ${\cal K}={\cal H}^*$ while $B_\mu ={\cal B}_\mu $ and $D={\cal D}$ are real:
\begin{equation}
  \left\{C,\,\zeta ,\,{\cal H},\, {\cal H}^*,\, B_\mu ,\,\lambda ,\, D\right\}\,.
 \label{realmultiplet}
\end{equation}

The complex multiplet reduces to a \emph{chiral multiplet} for $P_R{\cal Z}=0$. This then implies for consistency with (\ref{complexmult})
\begin{equation}
  P_R{\cal Z}=0\,,\qquad {\cal K}=0\,,\qquad {\cal B}_\mu =\rmi{\cal D}_\mu {\cal C}\,,\qquad \Lambda =0\,,\qquad {\cal D}=0\,.
 \label{chiralrestrcomplex}
\end{equation}
The remaining components can then be expressed in terms of the variables $\{X,P_L\V ,F\}$ of a chiral multiplet \cite{Kugo:1982cu}:
\begin{equation}
  \left\{X,\,-\rmi\sqrt{2}P_L\Omega,\,-2F,\,0,\,\rmi{\cal D}_\mu X,\,0,\,0\right\}\,.
 \label{chiralascomplex}
\end{equation}
However, for the conformal theory, this is only consistent when the chiral weight of ${\cal C}$ is equal to its Weyl weight: $(w,c)=(w,w)$.

Similarly for the antichiral multiplet $\{X^*,\, P_R\Omega ,\,F^*\}$, we have
\begin{equation}
  \left\{X^*,\,\rmi\sqrt{2}P_R\Omega,\,0,\,\,-2F^*,\,-\rmi{\cal D}_\mu X^*,\,0,\,0\right\}\,.
 \label{antichiralascomplex}
\end{equation}

\section{Multiplication laws}
The tensor calculus in conformal and Poincar\'{e} ${\cal N}=1$ supergravity was developed in several papers, the most complete version was given in  \cite{Kugo:1982cu, Kugo:1983mv}. Other most relevant papers are \cite{Ferrara:1978jt, *Ferrara:1978wj, Stelle:1978wj, *Cremmer:1978iv, *Cremmer:1978hn, *vanNieuwenhuizen:1978st, *Sohnius:1982xs, *Cremmer:1982wb, *Cremmer:1982en, VanProeyen:1979ks, VanProeyen:1983wk}.

The components of the multiplet $\tilde {\cal C}=f({\cal C}^i)$ are\footnote{Note that we use the Majorana conjugate, i.e. $\bar{\cal Z}={\cal Z}^T C $, which is for these complex spinors different from the Dirac conjugate.}
\begin{eqnarray}
 \tilde {\cal C} & = & f\,, \nonumber\\
 \tilde {\cal Z} & = & f_i{\cal Z}^i\,,\nonumber\\
 \tilde {\cal H} &=& f_i{\cal H}^i -\ft12f_{ij}\bar {\cal Z}^iP_L{\cal Z}^j\,,\nonumber\\
 \tilde {\cal K} &=& f_i{\cal K}^i -\ft12f_{ij}\bar {\cal Z}^iP_R{\cal Z}^j\,,\nonumber\\
 \tilde {\cal B}_\mu &=& f_i{\cal B}_\mu^i +\ft14\rmi f_{ij}\bar {\cal Z}^i\gamma _*\gamma _\mu {\cal Z}^j=f_i{\cal B}_\mu^i +\ft12\rmi f_{ij}\bar {\cal Z}^iP_L\gamma _\mu {\cal Z}^j\,,\nonumber\\
 \tilde {\Lambda }&=&f_i\Lambda ^i+\ft12 f_{ij}\left[\rmi\gamma _*\slashed{\cal B}^i+P_L{\cal K}^i+P_R{\cal H}^i-\slashed{\cal D}{\cal C}^i\right]{\cal Z}^j-\ft14f_{ijk}{\cal Z}^i\bar {\cal Z}^j{\cal Z}^k\,,\nonumber\\
 \tilde {\cal D}&=&f_i{\cal D}^i+\ft12f_{ij}\left({\cal K}^i{\cal H}^j-{\cal B}^i\cdot {\cal B}^j-{\cal D}{\cal C}^i\cdot {\cal D}{\cal C}^j-2\bar \Lambda ^i{\cal Z}^j-\bar {\cal Z}^i\slashed{\cal D}{\cal Z}^j\right)\nonumber\\
 &&-\ft14f_{ijk}\bar {\cal Z}^i\left(\rmi\gamma _*\slashed{\cal B}^j+P_L{\cal K}^j+P_R{\cal H}^j\right){\cal Z}^k +\ft18f_{ijk\ell}\bar {\cal Z}^iP_L{\cal Z}^j\bar {\cal Z}^kP_R{\cal Z}^{\ell}\,.
 \label{componentscompositecomplex}
\end{eqnarray}
In case the superfields have some spinor indices they may be treated as
 part of the indices $i$ in these formula. One can also use the explicit expressions in \cite{Kugo:1983mv}.
Note that
\begin{equation}
  P_Lf_{ijk}{\cal Z}^i\bar {\cal Z}^j{\cal Z}^k=f_{ijk}P_L{\cal Z}^i\bar {\cal Z}^jP_R{\cal Z}^k\,,
 \label{f3LR}
\end{equation}
due to the Fierz identity
\begin{equation}
  P_L{\cal Z}^{(i}\bar {\cal Z}^jP_L{\cal Z}^{k)}=0\,,
 \label{FierzsymmL}
\end{equation}
where we indicated the symmetric part in $(ijk)$.

If only chiral and antichiral multiplets occur, hence $f(X^\alpha ,\bar X^{\bar \alpha })$, for multiplets\\ $\{X^\alpha ,P_L\Omega ^\alpha ,F^\alpha \}$ and
$\{X^{\bar \alpha} ,P_R\Omega ^{\bar \alpha} ,F^{\bar \alpha} \}$ (and $\alpha $ and $\bar \alpha $ may run over a different range) this reduces to
\begin{eqnarray}
\tilde {\cal C}&=& f\,,\nonumber\\
 \tilde {\cal Z}&=&\rmi\sqrt{2}\left(- f_\alpha
\Omega^\alpha+ f_{\bar \alpha}
\Omega^{\bar \alpha}\right)\,, \nonumber\\
 \tilde {\cal H}&=& -2f_\alpha F^\alpha+f_{\alpha \beta }\bar
\Omega^\alpha  \Omega^\beta, \nonumber\\
 \tilde {\cal K}&=& -2f_{\bar \alpha} F^{\bar \alpha}+f_{\bar \alpha \bar \beta }\bar
\Omega^{\bar \alpha}  \Omega^{\bar \beta}, \nonumber\\
 \tilde {\cal B}_\mu&=&\rmi f_\alpha {\cal D} _\mu
X^\alpha-\rmi f_{\bar \alpha} \partial _\mu \bar
X^{\bar \alpha}+\rmi f_{\alpha \bar\beta }\bar
\Omega^\alpha \gamma _\mu\Omega^{\bar\beta},\nonumber\\
P_L \tilde {\Lambda }&=&-\sqrt{2}\rmi f_{\bar \alpha \beta
}\left[ (\slashed{\cal D } X^{ \beta } )\Omega^{\bar  \alpha}
-F^{\bar \alpha} \Omega^{\beta }\right]
-\frac{\rmi}{\sqrt{2}}f_{\bar \alpha\bar  \beta \gamma  }\Omega^{\gamma }\bar \Omega^{\bar\alpha}\Omega^{\bar\beta},\nonumber\\
P_R \tilde {\Lambda }&=&\sqrt{2}\rmi f_{\alpha \bar\beta
}\left[ (\slashed{\cal D }\bar X^{\bar \beta } )\Omega^\alpha
-F^\alpha \Omega^{\bar \beta }\right]
+\frac{\rmi}{\sqrt{2}}f_{\alpha \beta\bar \gamma  }\Omega^{\bar
\gamma }\bar \Omega^\alpha P_L\Omega^\beta,\nonumber\\
 \tilde {\cal D}&=& 2f_{\alpha \bar \beta }\left( -{\cal D} _\mu X^\alpha {\cal D}^\mu \bar X^{\bar \beta }-\ft12\bar \Omega^\alpha  P_L
  {\slashed{\cal D }}\Omega^{\bar \beta }-\ft12\bar \Omega^{\bar \beta }  P_R  {\slashed{\cal D }}\Omega^\alpha+F^\alpha  F^{\bar \beta }\right) \nonumber\\[3pt]
&&  f_{\alpha \beta \bar \gamma }\left( -\bar
\Omega^\alpha \Omega^\beta  F^{\bar \gamma }+\bar
\Omega^\alpha (\slashed{\cal D }X^\beta) \Omega^{\bar \gamma
}\right)  + f_{\bar \alpha \bar \beta  \gamma }\left( -\bar\Omega^{\bar \alpha}
 \Omega^{\bar \beta}  F^{ \gamma }+\bar\Omega^{\bar \alpha}
 (\slashed{\cal D }X^{\bar \beta})
\Omega^{ \gamma}\right)
\nonumber\\
&&+\,\ft12f_{\alpha \beta \bar \gamma \bar \delta }\,\bar
\Omega^\alpha P_L \Omega^\beta \bar \Omega^{\bar \gamma
}P_R\Omega^{\bar \delta }\,.
\label{DactionK}
\end{eqnarray}
We did not consider here chiral multiplets that transform under the gauge group, in which case a few extra terms appear \cite[Sec. 14.4.3]{Freedman:2012zz}.

On the other hand: if all multiplets are real multiplets $f(C^i)$ where the multiplets are of the form in (\ref{realmultiplet}) we get \cite{VanProeyen:1983wk}
\begin{eqnarray}
 \tilde {\cal C} & = & f\,, \nonumber\\
 \tilde {\cal Z} & = & f_i\zeta ^i\,,\nonumber\\
 \tilde {\cal H} &=& f_i{\cal H}^i -\ft12f_{ij}\bar \zeta^iP_L\zeta^j\,,\nonumber\\
 \tilde {\cal K} &=& f_i{\cal H}^{*i} -\ft12f_{ij}\bar \zeta^iP_R\zeta^j\,,\nonumber\\
 \tilde {\cal B}_\mu &=& f_iB_\mu^i +\ft14\rmi f_{ij}\bar \zeta^i\gamma _*\gamma _\mu \zeta^j \,,\nonumber\\
 \tilde {\Lambda }&=&f_i\lambda ^i+\ft12 f_{ij}\left[\rmi\gamma _*\slashed{B}^i+\Re{\cal H}^i-\rmi\gamma _*\Im{\cal H}^i-\slashed{\cal D}C^i\right]\zeta^j-\ft14f_{ijk}\zeta^i\bar \zeta^j\zeta^k\,,\nonumber\\
 \tilde {\cal D}&=&f_iD^i+\ft12f_{ij}\left({\cal H}^i{\cal H}^{*j}-B^i\cdot B^j-{\cal D}{\cal C}^i\cdot {\cal D}{\cal C}^j-2\bar \lambda ^i\zeta^j-\bar \zeta^i\slashed{\cal D}\zeta^j\right)\nonumber\\
 &&-\ft14f_{ijk}\bar \zeta^i\left(\rmi\gamma _*\slashed{B}^j+\Re{\cal H}^j-\rmi\gamma _*\Im{\cal H}^j\right)\zeta^k +\ft18f_{ijk\ell}\bar \zeta^iP_L\zeta^j\bar \zeta^kP_R\zeta^{\ell}\,.
 \label{componentscompositereal}
\end{eqnarray}

\section{Chiral projection for the complex multiplets}

The operation $T$ acts on a complex multiplet with $c=w-2$, producing a chiral multiplet with first component
\begin{equation}
  T({\cal C})= -\ft12{\cal K} \,.
 \label{TonC}
\end{equation}
This transforms left chiral according to (\ref{complexmult}).
The condition on the weights follows from the requirement that the Weyl and chiral weights of the components of ${\cal K}$ are equal.
In superspace $T$ is the operation $\bar D^2$ (see translation in Appendix \ref{app:superspace}).

It is useful to consider also $T({\cal C}^*)$, for which we then need that ${\cal C}$ has $c=2-w$. The first component is then
\begin{equation}
 T({\cal C}^*)=-\ft12 {\cal H}^*\,.
 \label{TonCstar}
\end{equation}
In summary
\begin{eqnarray}
  T({\cal C})&:\qquad &{\cal C}\ :\ (w,w-2)\,,\qquad T({\cal C})\ :\ (w+1, w+1)\nonumber\\
  T({\cal C}^*)&:\qquad &{\cal C}\ :\ (w,2-w)\,,\qquad
{\cal C}^*\ :\ (w,w-2)\,,\qquad T({\cal C}^*)\ :\ (w+1,w+1)\,.
  \label{SigmaTgeneral}
\end{eqnarray}
We can thus express this as "$T$ carries weights $(1,3)$". Further, note that in
\cite{Kugo:1983mv} this operation has also been defined for multiplets with external spinor indices, with the same restriction on weights, and
restriction to multiplets with only chiral external indices.

The components of $T({\cal C})$ are
\begin{equation}
 T({\cal C})= \left\{ -\ft12{\cal K}, -\ft12{\sqrt{2}}\,  \rmi \,P_L(\slashed{\cal D}{\cal Z}+\Lambda), \ft12({\cal D}+\bbox^C {\cal C}+\rmi\, {\cal D}_a {\cal B}^a )\right\}\,.
 \label{componentscalC}
\end{equation}

If ${\cal C}$ is a chiral multiplet, i.e. of the form (\ref{chiralascomplex}) then the condition on the weights for $T({\cal C})$ is not compatible with $w=c$. However, if this chiral multiplet has $w=1$, then  $T({\cal C}^*)$ is defined and is a chiral multiplet with $w=2$. This is then the map  \cite[(16.36)]{Freedman:2012zz} that associates to a chiral multiplet $\{X,\,P_L\Omega ,\, F\}$ the multiplet $\{X',\,P_L\Omega' ,\, F'\}$
\begin{equation}
 X'=F^*\,,\qquad P_L\Omega '=\slashed{\cal D}P_R\Omega \,, \qquad
 F'=\bbox^C  X^*\,.
 \label{chiralbarF}
\end{equation}

Note that an antisymmetric tensor multiplet (or `linear multiplet') is defined as a real multiplet with $T(C)=0$.

\section{Action formulae}
We will explain the notations for the actions: $[C]_D$ and $[X]_F$.
We start from the action formula of a chiral multiplet, which is \cite[(16.35)]{Freedman:2012zz}, where the notation $S_F$ is used. Here, and in several other places we use another notation: $[\cdots]_F$, the corresponding actions differ by a factor 2.
For a chiral multiplet $\{X,P_L\Omega ,F\}$, this is
\begin{equation}
  [X]_F = \int \rmd^4x\,e\left[ F +\frac{1}{\sqrt{2}}\bar \psi_\mu \gamma
  ^\mu P_L\Omega  + \frac12X \bar \psi _\mu \gamma ^{\mu \nu }P_R\psi
  _\nu\right] +\hc = \int \rmd^4x\,e\, 2 \Re F+\ldots \,.
\label{SF}
\end{equation}
Note that this is only applicable to a chiral multiplet with $w=3$, such that $F$ has weights $(4,0)$.

Then we define the action formula for a complex multiplet by
\begin{equation}
  [{\cal C}]_D= \ft12[T({\cal C})]_F\,.
 \label{DactionfromF}
\end{equation}
This is only consistent if the weights of ${\cal C}$ are $(2,0)$.
The fact that the last term in (\ref{componentscalC}) is a total derivative, leads to the equation
\begin{equation}
  \left[ T({\cal C})\right]_F = \left[ T({\cal C}^*)\right]_F\,,
 \label{thmimaginary}
\end{equation}
when both sides of the equation are defined in the conformal setting, i.e. ${\cal C}$ should have $(w,c)=(2,0)$.

Therefore, the action depends in fact on the real part of the multiplet.
For a real multiplet with Weyl weight $w=2$ \cite{Stelle:1978wj,Kugo:1982cu}:
\begin{equation}
  [C]_D  = \frac12 \int \rmd^4x\,e\left[D + {\cal D}^a{\cal D}_a C +\ft12\left( \rmi\bar \psi\cdot \gamma
 P_R (\lambda +\slashed{\cal D}\zeta)-\ft14\bar \psi _\mu P_L\gamma ^{\mu \nu }\psi _\nu {\cal H}+\hc\right)\right]\,.
\label{SDconf1}
\end{equation}

After using expressions for the dependent superconformal gauge fields and further
manipulations discussed in \cite[Appendix 17B]{Freedman:2012zz} it can be written as
\begin{eqnarray}
 [C]_D  &=& \frac12 \int \rmd^4x\,e\left[D -\ft12\bar \psi\cdot \gamma\rmi\gamma _*\lambda
     -\ft13C\,R(\omega )+\ft16\left( C\,\bar \psi_\mu \gamma ^{\mu \rho \sigma }- \rmi\bar\zeta  \gamma ^{\rho \sigma }\gamma _*\right)
 R'_{\rho \sigma }(Q)
\right.\nonumber\\
   &&\,\phantom{\frac12\int \rmd^4x\,e[}\left.
   +\,\ft14\varepsilon ^{abcd}\bar \psi _a\gamma _b\psi _c\left( B_d-\ft12\bar \psi _d\zeta \right) \right] \,.
\label{SDconf}
\end{eqnarray}

\subsection{Theorems on \texorpdfstring{$T$}{T}-operation}

The $T$-operation vanishes on a chiral multiplet, and moreover
\begin{equation}
  T(Z{\cal C})= Z\,T({\cal C})\,,
 \label{chiralTzero}
\end{equation}
if $Z$ is a chiral multiplet and the weights of ${\cal C}$ satisfy the first condition in (\ref{SigmaTgeneral}).
This theorem follows directly from the expression of $\tilde {\cal K}$ in (\ref{DactionK}) applied to $Z{\cal C}$. Indeed, the $T$-operation consists in (up to normalization) taking the chiral multiplet defined by the ${\cal K}$ component.
Since in the chiral multiplet the ${\cal K}$ component and the right-handed component of ${\cal Z}$ vanish, the only remaining term is the product of the lowest component $Z$ and the ${\cal K}$ component of ${\cal C}$.

Observe that this theorem remains true if the multiplets are spinor multiplets. The extra spinor indices are then external indices, playing the same role as the indices $i,j$ in the product rules  (\ref{componentscompositecomplex}).

One application of these equations is the following theorem \cite{Cecotti:1987sa}.
For any two chiral multiplets $\Lambda $ (with $w=0$) and $Z$ (with $w=1$) we have
\begin{equation}
 [(\Lambda +\Lambda ^*)ZZ^*]_D=  [\Lambda Z T (Z^*)]_F\,.
 \label{thmDtoF}
\end{equation}
To prove this, the first step is to translate the left-hand side to the chiral form:
\begin{equation}
 [(\Lambda + \Lambda^* )Z  Z^*]_D = \ft12\left[T\left((\Lambda + \Lambda^* )Z  Z^*\right)\right]_F\,.
 \label{tochiralF}
\end{equation}
Then (\ref{thmimaginary}) implies that the  term with $\Lambda ^*$  is equal to the one with $\Lambda $, and  (\ref{chiralTzero}) implies that $T$ acts only on $ Z^*$. This proves (\ref{thmDtoF}).

\section{Solution of  EOM for \texorpdfstring{$\mathbf{\cal V}$}{V} versus constraints}\label{sec:EOMvsConstraints}
It was shown in \cite{Bergshoeff:2016psz} that the equations of motion before or after the constraints are equivalent. This means the following:
We start with the  action with Lagrange multipliers $\lambda^i$, fields $X^\alpha$ and  $F^a$:
\begin{equation}
  S(F^a,X^\alpha,\lambda^i)=S_1(F^a,X^\alpha) + \lambda^i C_i(F^a,X^\alpha)\,.
 \label{LwithLagrmult}
\end{equation}
Assume that
\begin{equation}
  X^\alpha =x^\alpha(F^a)\,,
 \label{solveconstraints}
\end{equation}
 solves the constraints, i.e.
\begin{equation}
  C_i(F^a,x^\alpha(F^a))=0\,.
 \label{Csolved}
\end{equation}
In such a case, it was shown in \cite{Bergshoeff:2016psz} that such a solution of the constraints, solving also the other field equations of (\ref{LwithLagrmult}), should be a solution of the effective Lagrangian where the constraints are already inserted:
\begin{equation}
  S_{\rm eff} (F^a)= S_1(F^a,x^\alpha(F^a))\,.
 \label{Lsolved}
\end{equation}
This solves  in general a problem raised in \cite[Appendix C]{Kuzenko:2011tj}.

We now prove that for any function of superfields $f(C^i)$ it is equivalent to impose the constraint $f(C^i)=0$ or to add the Lagrange multiplier $[\V f(C^i)]_D$ to the action and solve the equations of motion for the components of the complex multiplet $\V$. Since $[\V f(C^i)]_D$ is linear in the components of $\mathbf{\cal V }$ and linear in the components of $f(C^i)$, we find trivially that $f(C^i)=0$ solves the equations of motion arising from the LM term $[\V f(C^i)]_D$. However, since the equations of motion are linear combinations it is not immediately obvious that $f(C^i)=0$ is the unique solution to the LM equations of motions.

We start with the action for our LM term: $[\V f(C^i)]_D$
\ba\label{eq:action}
[\V f(C^i)]_D &=&\frac14 \int {\rm d}^4x\,e\left[\tD -\frac12\bar \psi\cdot \gamma\rmi\gamma _* \Lambda  -\frac 13 \tC \,R(\omega )+\frac16\left( \tC\,\bar \psi_\mu \gamma ^{\mu \rho \sigma }- \rmi\bar{{\cal Z}}  \gamma ^{\rho \sigma}\gamma _*\right)  R' _{\rho \sigma }(Q)\right.\cr
&&\,\phantom{\frac12\int {\rm d}^4x\,e[} \left. +\,\frac14\varepsilon ^{abcd}\bar \psi _a\gamma _b\psi _c\left(\tilde{\mathcal{B}}_d-\frac12\bar \psi _d{\cal Z} \right) \right] +\hc\,.
\ea
Denoting the components as follows
\be
\begin{array}{ccccccccccc}\label{eq:table}
\V & = & \{ & \C, & \Z, & \H, & \K, & \B, & \L, & \D & \}\,,\\
f({\cal C}^i) &=&   \{ & \C^f, & \Z^f, & \H^f, & \K^f, & \B^f, & \L^f, & \D^f & \}\,,\\
\V f({\cal C}^i) &=&   \{ & \tC, & \tZ, & \tH, & \tK, & \tB, & \tL, & \tD & \}\,,
\end{array}
\ee
we find from the multiplication rules in \eqref{componentscompositecomplex}
\ba
\tilde{\C} &=& \C \C^{f} \,,\cr
\tZ &=& \C^{f} \Z +\C \Z^{f}\,,\cr
\tH &=& \H\C^{f} +\C \H^{f}- \overline{\Z}^{f} P_L \Z \,,\cr
\tK &=& \K \C^{f} +\C \K^{f}- \overline{\Z}^{f} P_R \Z  \,,\cr
\tB &=&\B \C^{f}+\C \B^{f}+\frac{\rmi}{2}\overline{\Z}^{f} \gamma _* \gamma_\mu \Z\,,\cr
\tL &=&  \C^{f} \L+\C \L^{f} +\frac12 \lp \rmi\gamma_* \slashed{\mathcal{B}}+P_L \K+P_R \H -\slashed{\D}\C \rp\Z^{f} \cr
&&+\frac12 \lp \rmi \gamma_* \slashed{\mathcal{B}}^{f}+ P_L \K^{f} + P_R \H^{f}-\slashed{\D} \C^{f} \rp \Z \,,\cr
\tD &=& \D \C^{f}+\C\D^{f} +\frac12 \K \H^{f}+\frac12 \H \K^{f}-\B^{f} \mathcal{B}^\mu-\bar{\L}^{f} \Z-\overline{\Z}^{f} \L-(\D^\mu \C) (\D_\mu \C^{f}) \cr
&&-\frac12 \overline{{\cal Z}}  \slashed{\D} \Z^{f} -\frac12\overline{\Z}^{f}  \slashed{\D} \Z \,.
\label{eq:Cf}
\ea
One can see that the Lagrange multiplier ${\cal D}$ appears only in the component $\tilde {\cal D}$. Therefore its equation of motion is simply that ${\cal C}^f=0$.
Therefore, the equations of motion of all the Lagrange multipliers imply that all components of $f({\cal C}^i)$  vanish, as we now show.

Consider in general an invariant action
\begin{equation}
  S= L_i\, C^i\,, \qquad (\delta L_i)\, C^i + L_i\,\delta  C^i=0\,.
 \label{SinvariantLC}
\end{equation}
Here $L_i$ are some Lagrange multipliers that do not appear in other terms of the action, and $C^i$ are the constraints that then follow from the equations of motion of $L_i$. We use the Bryce deWitt notation in which the sum over the indices $i$ contains a spacetime integral. Now its clear that for any solution of the constraint, $C^i=0$, we should have that $\delta C^i=0$. Hence, if we have solved one particular constraint, then everything that follows from taking transformations should also be true when all equations of motion of the Lagrange multipliers are satisfied. At the end one checks that with the equations that one has obtained, all $C^i$ vanish.

It can be nontrivial to know which initial equation is sufficient to get all the equations that one needs. One has to look for an equation of low dimension, and supersymmetry gives then constraints of higher dimension. E.g. in the case above, we obtain ${\cal C}^f=0$. Then the transformation of this gives ${\cal Z}^f=0$, and so on for all the components. Hence all constraints are solved.

\section{Example of orthogonal nilpotent multiplets}

If we want to treat a $D$-term as in \cite[(3.11)]{Ferrara:2015tyn}, we have to assign the weights as follows:
In order to be consistent with eq. (1.2) in that paper, and $\Phi $ is a chiral multiplet, $\mathbf{\Phi} ,\, \overline{\mathbf{\Phi} }$ and $\mathbf{B}$ should have weights $(0,0)$. Assigning weights $(w_S,w_S)$ to $\mathbf{S}$, the weights of the LM $\mathbf{\V }$ should be $(2-w_S, -w_S)$ in order to construct such a $D$-term. $w_S$ can be chosen conveniently. Then the weights of the chiral $\mathbf{\Lambda }$ in (3.8) should be $(3-2w_S, 3-2w_S)$.

If one associates $S$ to one of the fields $X^I/X^0$ in the usual conformal approach, then $w_S=0$.

\subsection{The Lagrange multiplier \texorpdfstring{$[{\bf \V S B}]_D$}{[Omega S B]D}}
We take $\mathbf{\cal V }$ to be a complex multiplet, ${\bf S}$ to be a (nilpotent) chiral multiplet and ${\bf B}=\frac{1}{2\rmi}({\bf \Phi -\bar \Phi})$ to be a real multiplet. We use the notation
\be
\begin{array}{ccccccccccc}
{\mathbf{\cal V }} & = & \{ & \C, & \Z, & \H, & \K, & \B, & \L, & \D & \}\,,\\
{\bf S} &=&   \{ &s,&-\rmi \sqrt{2}P_L\Omega^S,&-2F^S,&0,& \rmi \D_\mu s,&0,&0&\}\,,\\
{\bf \Phi} &=&   \{& \vp+\rmi b,&-\rmi \sqrt{2}P_L\Op,&-2F^\Phi,&0,& \rmi \D_\mu (\vp+\rmi b),&0,&0&\}\,,\\
{\bf \bar{\Phi}} &=&   \{ &\vp-\rmi b,&\rmi \sqrt{2}P_R\Op,&0,&-2F^{\Phi*},&- \rmi \D_\mu (\vp-\rmi b),&0,&0&\}\,,\\
{\bf B} &=&   \{& b,&-\frac{1}{\sqrt{2}} \Op,& \rmi F^\Phi,& -\rmi F^{\Phi*},&\D_\mu \vp,&0,&0&\}\,,
\label{table}\end{array}
\ee
where $b$ and $\vp$ are real and all other quantities are complex. Using the multiplication formulas, we find the following components for ${\bf S \bar{\Phi}} $
\ba
\C^{S\bar\Phi} &=& s (\vp -\rmi b)\,,\cr
\Z^{S\bar\Phi} &=& -\rmi \sqrt{2} (\vp-\rmi b) P_L\Omega^S + \rmi \sqrt{2} s P_R\Op  \,, \cr
\H^{S\bar\Phi} &=&-2F^S (\vp-\rmi b)  \,, \cr
\K^{S\bar\Phi} &=& -2 s F^{\Phi*} \,, \cr
\B^{S\bar\Phi} &=& \rmi(\vp-\rmi b) \D_\mu s-\rmi s \D_\mu (\vp-\rmi b) + \rmi \bar{\O}^S P_L \gamma_\mu \Op \,, \cr
\L^{S\bar\Phi} &=&-\rmi\sqrt{2} \lp F^S+\slashed{\D} s\rp P_R\Op + \rmi \sqrt{2} \lp F^{\Phi*}+\slashed{\D}(\vp-ib)\rp P_L \O^S\,, \cr
\D^{S\bar\Phi} &=& 2 F^S F^{\Phi*} - 2 (\D^\mu (\vp-\rmi b))(\D_\mu s)-\bar{\Omega}^S P_L  \slashed{\D} \Op - \bar{\Omega}^\Phi \slashed{\D} P_L \O^S  \,.\quad
\ea

We can also calculate the components of ${\bf S B}$
\ba\label{eq:SB}
\C^{SB} &=& s b\,,\cr
\Z^{SB} &=& -\rmi b \sqrt{2}P_L\Omega^S - \frac{1}{\sqrt{2}} s \Op \,, \cr
\H^{SB} &=&-2F^S b +\rmi s F^\Phi-\rmi\bar{\O}^S P_L \Op \,, \cr
\K^{SB} &=& -\rmi s F^{\Phi*} \,, \cr
\B^{SB} &=& \rmi b \D_\mu s+s \D_\mu \vp- \frac12 \bar{\O}^S P_L \gamma_\mu \Op \,, \cr
\L^{SB} &=&\frac{1}{\sqrt{2}} \lp F^S+(\slashed{\D} s) \rp P_R \Op -\frac{1}{\sqrt{2}} \lp\slashed{\D} (\vp-\rmi b) + F^{\Phi*} \rp P_L \O^S\,, \cr
\D^{SB} &=&\rmi F^S F^{\Phi*} - \rmi (\D^\mu (\vp-\rmi b))(\D_\mu s)-\frac{\rmi}{2} \bar{\Omega}^S P_L \slashed{\D} \Op - \frac{\rmi}{2} \bar{\Omega}^\Phi \slashed{\D} P_L \O^S  \,.\quad
\ea

Similarly, we find the following components for ${\bf \tilde C} = {\bf \V S B}$
\ba
\tilde{\C} &=& \C s b \,,\cr
\tZ &=& \Z s b-\rmi \sqrt{2} \C b P_L \O^S - \frac{1}{\sqrt{2}} \C s \Op\,,\cr
\tH &=& \H s b -2 \C F^S b +\rmi \C s F^\Phi + \rmi \sqrt{2} b \bar{\Omega}^S P_L \Z +\frac{1}{\sqrt{2}} s \bar{\O}^\Phi P_L \Z - \rmi \C \bar{\O}^S P_L \Op\,,\cr
\tK &=& \K s b -\rmi \C s F^{\Phi*}  +\frac{1}{\sqrt{2}} s \bar{\O}^\Phi P_R \Z \,,\cr
\tB &=&\B s b+\rmi \C b \D_\mu s+\C s \D_\mu \vp +\frac{1}{\sqrt{2}} b \bar{\O}^S P_L \gamma_\mu \Z +\frac{\rmi}{2\sqrt{2}}s \bar{\O}^\Phi \gamma_\mu \gamma_* \Z-\frac{1}{2} \C \bar{\O}^S P_L \gamma_\mu \Op\,,\cr
\tL &=& \L s b -\frac{\rmi}{\sqrt{2}} b \lp \rmi \gamma_* \slashed{\mathcal{B}} + \K -\slashed{\D} \C \rp P_L \O^S -\frac{1}{2\sqrt{2}} s \lp \rmi \gamma_* \slashed{\mathcal{B}} + \K P_L+\H P_R -\slashed{\D} \C \rp \Op\cr
&& - b \lp \slashed{\D}s + F^S\rp P_R\Z -\frac{\rmi}{2} s\lp  \slashed{\D} (\vp \gamma_* -\rmi b) + F^{\Phi*} P_L - F^\Phi P_R \rp \Z\cr
&& -\frac{1}{\sqrt{2}} \C\lp  \slashed{\D} (\vp -\rmi b) + F^{\Phi*} \rp P_L \O^S +\frac{1}{\sqrt{2}} \C \lp \slashed{\D}s+ F^S\rp P_R \Op\cr
&&-\frac{\rmi}{2} \Z\ \bar{\O}^\Phi P_L \O^S -\frac{\rmi}{2} P_L \O^S \bar{\O}^\Phi \Z-\frac{\rmi}{2} \Op \bar{\O}^S P_L \Z\,,\cr
\tD &=& \D s b \cr
&&+ b \lp - \K F^S - \rmi \B (\D^\mu s) - (\D_\mu \C)(\D^\mu s)  \right.\nonumber\\
&&\phantom{+ b }\left.+ \rmi \sqrt{2} \bar{\O}^S P_L \L +\frac{\rmi}{\sqrt{2}} \overline{{\cal Z}} \slashed{\D}P_L\O^S  + \frac{\rmi}{\sqrt{2}} \bar{\O}^S P_L \slashed{\D} \Z \rp \cr
&&+\frac{s}{2} \lp\rmi F^\Phi \K-\rmi F^{\Phi*} \H -2 \B \D^\mu \vp -2 (\D_\mu \C) (\D^\mu b)+\sqrt{2} \bar{\Omega}^\Phi \L \right.\nonumber\\
&&\phantom{+\frac{s}{2}}\left.
+\frac{1}{\sqrt{2}} \bar{\cal Z}{\slashed{\D} \Op}  +\frac{1}{\sqrt{2}} \bar{\O}^\Phi \slashed{D} \Z\rp\label{D}\\
&&+\C \lp\rmi F^S F^{\Phi*} -\rmi (\D^\mu (\vp-\rmi b))(\D_\mu s) -\frac{\rmi}{2} \bar{\O}^\Phi \slashed{\D} P_L \O^S -\frac{\rmi}{2} \bar{\O}^S P_L \slashed{\D} \Op\rp\cr
&& +\frac{1}{2\sqrt{2}} \bar{\O}^\Phi (\slashed{D}s\gamma _* - 2F^S P_R)\Z - \frac{1}{\sqrt{2}} \bar{\O}^S P_L( \slashed{\D} \vp \ -F^{\Phi*})  \Z+\frac{1}{2}  \bar{\O}^S P_L
( \slashed{\mathcal{B}} - \rmi \K) \Op \,.\nonumber
\ea
Using the result for ${\bf SB}$ in \eqref{eq:SB} we can rewrite this as
\ba
\tilde{\C} &=& \C \C^{SB} \,,\cr
\tZ &=& \Z \C^{SB}+\C \Z^{SB}\,,\cr
\tH &=& \H\C^{SB} - \bar{\Z}^{SB} P_L \Z +\C \H^{SB}\,,\cr
\tK &=& \K \C^{SB}- \bar{\Z}^{SB} P_R \Z +\C \K^{SB} \,,\cr
\tB &=&\B \C^{SB}-\frac{\rmi}{2}\bar{\Z}^{SB} \gamma_\mu \gamma_* \Z+\C \B^{SB}\,,\cr
\tL &=& \L \C^{SB} +\frac12 \lp \rmi\gamma_* \slashed{\mathcal{B}}+\K P_L+\H P_R -\slashed{\D}\C \rp\Z^{SB} \cr
&&+\frac12 \lp \rmi \gamma_* \slashed{\mathcal{B}}^{SB}+ \K^{SB} P_L + \H^{SB} P_R-\slashed{\D} \C^{SB} \rp \Z +\C \L^{SB}\,,\cr
\tD &=& \D \C^{SB}-\bar{\Z}^{SB} \L +\frac12 \K \H^{SB}+\frac12 \H \K^{SB}-\B^{SB} \mathcal{B}^\mu-\bar{\L}^{SB} \Z+\C\D^{SB}-(\D^\mu \C) (\D_\mu \C^{SB}) \cr
&&-\frac12 \overline{\Z} \slashed{\D} \Z^{SB}  -\frac12 \overline{\Z^{SB}} \slashed{\D} \Z \,.
\label{D2}
\ea
This example illustrates the general point of Sec. \ref{sec:EOMvsConstraints} and demonstrates that once we solve the  equations of motion for the components of the general complex superfield $\mathbf{\cal V }$ we find that all components of the superfield ${\bf  S B}$ must vanish.

\section{Constraint that a multiplet is chiral, `relaxed constraint'}
Another constraint that has been considered is imposing that a multiplet is chiral or for the real case: that it is the sum of a chiral multiplet and its antichiral complex conjugate.
We will consider now this real case. We have a real multiplet  (\ref{realmultiplet}) and want to impose that it is the sum of a chiral and antichiral multiplet as given in (\ref{chiralascomplex}) and (\ref{antichiralascomplex}). Defining
\begin{equation}
  X= \frac{1}{\sqrt{2}}(A+\rmi B)\,,
 \label{XAB}
\end{equation}
the multiplet should thus be of the form
\begin{equation}
  \left\{\sqrt{2}A,\,-\rmi\sqrt{2}\gamma _*\Omega ,\,-2F,-2F^*,-\sqrt{2}{\cal D}_\mu B,0,0\right\}\,,\qquad {\cal D}_\mu B=\partial _\mu B+\ft12\rmi\bar \psi _\mu \gamma _*\Omega \,,
 \label{chiralplusanti}
\end{equation}
and it should have weights $(0,0)$ to be consistent with chiral $+$ antichiral.
Comparing to a general form of such a real multiplet, (\ref{realmultiplet}), the main constraint is $\lambda =0$. Then the form of (\ref{chiralplusanti}) can be shown to follow by supersymmetry.

A  direct approach to describing the models with `relaxed constraint'  is to use the \emph{linear multiplet} $L$ for the  LM superfield:
it is a real multiplet that satisfies
\begin{equation}
  \mbox{linear multiplet: }\quad  T(L)=0\,.
 \label{deflinmult}
\end{equation}
Hence, in order to be well defined, $L$ should have weights (2,0) \cite{deWit:1981fh}. Putting (\ref{componentscalC}) to zero, leads to a multiplet with components $\{L,\,\chi  ,\, E_{\mu \nu }\}$ where $L$ is real, $\chi $ is Majorana, and $E_{\mu \nu }$ is a gauge antisymmetric tensor. It is embedded in the complex multiplet in the form \cite{Kugo:1982cu}
\begin{equation}
  \left\{L,\,\chi  ,\, 0,\,0,\, E_a\,, -\slashed{\cal D}\chi  ,-\bbox^C L\right\}\,,\qquad {\cal D}_aE^a=0\,,
 \label{linmultcomp}
\end{equation}
where the last equation implies that $E^a$ is a covariant field strength of an antisymmetric tensor, $E_{\mu \nu }$ \cite{deWit:1981fh}:
\begin{equation}
  E^\mu =e^{-1}\varepsilon ^{\mu \nu \rho \sigma }\left(\partial _\nu E_{\rho \sigma }-\ft14\bar \psi _\nu \gamma _\rho \psi _\sigma L\right)+\bar \chi \gamma^{\mu \nu }\psi _\nu \,.
 \label{Bmusolved}
\end{equation}
The multiplet can also be written in terms of a chiral spinor prepotential $L_\alpha $:
\begin{equation}
  L={\cal D}^\alpha L_\alpha +\overline{{\cal D}}^{\dot \alpha }L_{\dot \alpha }\,.
 \label{Lasspinor}
\end{equation}
The weight requirements on spinor multiplets are discussed in Appendix \ref{app:GeneralMult}. They imply here that $L_\alpha $ has weights $(\ft32,\ft32,\ft12,0)$. The expression in (\ref{Lasspinor}) satisfies  (\ref{deflinmult}) using \cite[(3.22)]{Kugo:1983mv} and \cite[(3.9)]{Kugo:1983mv}.

We will now prove the following result, already mentioned in \cite[(3.18)]{Ferrara:1983dh}. If there is a linear multiplet $L$ that is used as a Lagrange multiplier with a term in the action of the form $[LU]_D$, where $U$ is real (and has Weyl weight~0), then the equations of motion of $L$ imply that $U$ is of the form (\ref{chiralplusanti}), i.e. a chiral + antichiral multiplet.

First, we prove that if $L$ is linear and $U= X+X^*$, where $X$ is chiral, then $[LU]_D=0$. Indeed, starting with (\ref{DactionfromF}):
\begin{equation}
  2\left[(X+X^*)L\right]_D=\left[T((X+X^*)L)\right]_F= 2\left[T(X\,L)\right]_F= 2\left[X\,T(L)\right]_F=0\,,
 \label{proofchirallin0}
\end{equation}
using in the different steps (\ref{thmimaginary}),  (\ref{chiralTzero}) and  (\ref{deflinmult}).

Using the prepotential formulation we can also write
\begin{equation}
  \left[LU\right]_D= \left[U\left({\cal D}^\alpha L_\alpha +\overline{{\cal D}}^{\dot \alpha }L_{\dot \alpha }\right)\right]_D\,.
 \label{LUDprepotential}
\end{equation}
If $L_\alpha $ is chiral, then its field equation is $T{\cal D}_\alpha V=0$. In order to work with primaries we should have
\begin{equation}
 L_\alpha \ :\ (\ft32,\ft32,\ft12,0)\,,\qquad U\ :\ (0,0,0,0)\,,\qquad {\cal D}^\alpha L_\alpha\ :\ (2,0,0,0)\,,\qquad D_\alpha U\ :\ (\ft12, -\ft32,\ft12,0)\,.
 \label{weightsWV}
\end{equation}
The main step will be to prove that
\begin{equation}
  \left[U{\cal D}^\alpha L_\alpha\right]_D =  \left[ L_\alpha{\cal D}^\alpha U\right]_D\,.
 \label{mainsteppdD}
\end{equation}
This follows from the fact that ${\cal D}_\alpha $ by its definition is distributive and
\begin{equation}
  \left[{\cal D}_\alpha V^\alpha \right]_D = 0\,.
 \label{totalderiv}
\end{equation}
Then we write
\begin{equation}
  \left[L \, U\right]_D = 2\left[({\cal D}^\alpha L _\alpha) U\right]_D = 2\left[ L _\alpha {\cal D}^\alpha U\right]_D = \left[T(L _\alpha {\cal D}^\alpha U)\right]_F\,.
 \label{stepsLU}
\end{equation}
Then, using that  (\ref{chiralTzero}) also holds for spinor multiplets, the field equations of $L _\alpha $ imply that $T({\cal D}^\alpha U)=0$. Indeed, in the multiplication of chiral multiplets, nor in the action formula appear derivatives on the fields, and the field equations are thus linear. $T({\cal D}^\alpha U)=0$ is the multiplet that starts with $\lambda $. Thus this is the condition that $U$ is of the form (\ref{chiralplusanti}): chiral + antichiral.

One more possibility to describe the models with `relaxed constraints' is the following.
In \cite{deWit:1981fh} the action for the product of the linear multiplet and another real multiplet,
\begin{equation}
  U=\left\{C,\zeta ,{\cal H},{\cal K},B_a,\lambda,D \right\}\,,
 \label{Ucomponents}
\end{equation}
has been obtained:
\begin{equation}
 \left[L\,U\right]_D=\ft12\int \rmd^4x\,e\left[L\, D -\bar \chi  \lambda -\ft12\rmi L\,\bar \psi_\mu \gamma ^\mu \gamma _*\lambda \right] -\varepsilon ^{\mu \nu \rho \sigma }V_\mu \partial _\nu E_{\rho \sigma }\,, \qquad V_\mu =B_\mu -\ft12\bar \psi _\mu \zeta \,.
 \label{LCD}
\end{equation}
This is in agreement with the calculation in (\ref{proofchirallin0}): it says that the only components of the real multiplet that appear are those that are invariant under $C\rightarrow C + \Lambda +\bar \Lambda $. Here one can see immediately that the field equations are
\begin{equation}
  D=\lambda =0\,,\qquad V_\mu =\partial _\mu  B\ \rightarrow \ B_\mu = -\sqrt{2}\partial _\mu  B -\ft12\bar \psi _\mu \zeta =-\sqrt{2}{\cal D}_\mu  B\,,
 \label{fecomponents}
\end{equation}
where $B$ is a priori an arbitrary field, which can be identified with the one in (\ref{chiralplusanti}), where $\zeta =-\rmi\sqrt{2}\gamma _*\Omega $.

\section{Summary}
The action \rf{Dterm}  has a  linearly realized supersymmetry and in addition to the physical superfields there are  Lagrange multiplier superfields $\Lambda^k$ and $\V^\ell$, all superfields are unconstrained. Integrating the LM's out from the action \rf{Dterm}
 we derive  the action with constraints imposed and a non-linearly realized supersymmetry. This is a property of the action \rf{Dterm} after the equations of motion for all LM are solved:
\begin{eqnarray}
{\cal L} ^{non-lin}&= &[N (X,\bar X)]_D + [\mathcal{W}(X)]_F +  [f_{AB} (X) \bar \lambda ^A P_L \lambda^B]_F+ \left[\sum _k \Lambda^k  A_k(X )\right] _F\nonumber\\ &+&\left[\sum _\ell \V^\ell  B_\ell(X, \bar X, \lambda^A,\cdots ) \right] _D \Bigg |_{ \frac{\delta {\cal L}^{lin}}{\delta \Lambda^k}=0, \, \frac{\delta {\cal L}^{lin}}{\delta \V^\ell}=0}\,.
\label{Dterm1}
\end{eqnarray}
But since we also have an underlying model with linear supersymmetry \rf{Dterm} where the  LM superfields are off shell, our  non-linear action \rf{Dterm1} follows from the linear one. Likewise the corresponding non-linear supersymmetry transformations can be deduced from the linear ones. A clear example of this was given in the case of one nilpotent superfield and one chiral LM superfield in  \cite{Ferrara:2014kva} which led to the explicit action of `pure de Sitter supergravity' in
 \cite{Bergshoeff:2015tra, *Hasegawa:2015bza}.

 Now we have a  general understanding of the relation between linear and non-linear models with LM superfields in F-terms and in D-terms: the non-linear models arise in our approach as  models where the equations of motion for the LM superfields have been solved. This explains why they are consistent and how the non-linear supersymmetry transformations rules follow from the linear ones. Since now our LM's are of a general nature, not constrained to be chiral, our new analysis includes the constrained superfields, like the orthogonal nilpotent superfields and the relaxed version of it, studied in the global supersymmetry case in \cite{Komargodski:2009rz,Kahn:2015mla} and developed to a local supersymmetry in \cite{Ferrara:2015tyn,Dall'Agata:2015lek}. Here we presented the underlying linear superconformal models for the  constrained non-linear models, which are of a particular  interest to cosmology \cite{Carrasco:2015iij}.

\section*{Acknowledgments}
We are grateful to I. Antoniadis, E. Bergshoeff, J. J. Carrasco,  G. Dall'Agata, E. Dudas, D. Freedman,  A. Karlsson, A. Kehagias, S. Kuzenko, A. Linde, B. Mosk, D. Murli, M. Porrati,  A. Sagnotti, J. Thaler and F. Zwirner for useful discussions and collaboration on related projects. The work of SF is supported in part by INFN-CSN4-GSS. The work of RK is
supported by the SITP, and by the NSF Grant PHY-1316699. AVP is supported in part by the Interuniversity Attraction Poles Programme initiated by the Belgian Science Policy (P7/37). This work has been supported in part by COST Action MP1210 `The String Theory Universe'. AVP and TW thank the Department of Physics of Stanford University for the hospitality during a visit in which this work was initiated.

\appendix
\section{Translation to superspace}
\label{app:superspace}

The recent work on component versus superspace approaches to conformal supergravity is in \cite{Kugo:2016zzf}.
Here we will use the translations to superspace as in \cite[Appendix 14A]{Freedman:2012zz}. We use the shortcuts
\begin{equation}
  D^2 = -\bar D P_L D \,,\qquad \bar D^2 = -\bar D P_R D\,,
 \label{D2barD2}
\end{equation}
where the bar in the right-hand sides is the Majorana bar. With spinor indices, $D^2= -D^\alpha (P_L)_\alpha{}^\beta D_\beta $ for $\alpha =1,\ldots ,4$ or in the 2-component notation:
$D^2= -D^\alpha D_\alpha =D_\alpha D^\alpha $.
It satisfies
\begin{equation}
 \left. D^2 \bar \theta P_L\theta \right|_{\theta =0}= -4\,.
 \label{D2theta2}
\end{equation}

Comparing with (\ref{SF}) and  (\ref{SDconf1}), we have
\begin{equation}
  [C]_D=\int \rmd^4x\, D^2 \bar D^2 C\,,\qquad [X]_F =\int \rmd^4x\, D^2 X +\hc\,.
 \label{DFsuperspace}
\end{equation}
To facilitate the writing we will define $\int \rmd^2\theta $ and $\int \rmd^4\theta $ as
\begin{equation}
 \int \rmd^2\theta = D^2\,,\qquad  \int \rmd^4\theta = \int \rmd^2\theta \,\rmd^2\bar \theta =D^2\bar D^2\,,
 \label{integraltheta}
\end{equation}
such that (identifying superfields by their first components)
\begin{equation}
  [C]_D =\int \rmd^4 x\,\rmd^4 \theta C\,, \qquad [X]_F= \int \rmd^4 x\,\rmd^2 \theta\, X +\hc\,.
 \label{DFinttheta}
\end{equation}
For global supersymmetry, the complex superfield is of the form
\begin{equation}
  \Phi ={\cal C}+\ft12\rmi\bar \theta \gamma _*{\cal Z}-\ft18\bar \theta P_L\theta {\cal H}-\ft18\bar \theta P_R\theta {\cal K}+\ldots \,.
 \label{complexsuperfield}
\end{equation}


\section{General multiplets}
\label{app:GeneralMult}

Now we consider general multiplets that have spin. We denote the spin as $(J_1,J_2)$, where e.g. a scalar multiplet has $(0,0)$, and a multiplet like $W_\alpha $ has $(\ft12,0)$. We use here the indices $\alpha $ for left-handed projections, and $\dot \alpha $ for right-handed projections of spinors.

There are a few ways to constrain multiplets and to build multiplets from other multiplets. The ${\cal D}_\alpha $ operation consists in taking the $P_L{\cal Z}$ component of the multiplet ${\cal C}$ as first component, while  $\overline{{\cal D}}_{\dot \alpha} $ takes the $P_R{\cal Z}$ component. The $T$ operation is defined in (\ref{TonC}) from the ${\cal K}$ component. However, we require then that these do not transform under $S$-supersymmetry, i.e. they are primary superconformal fields.
Here is a table of requirements on the weights in order that these objects are primary \cite{Kugo:1983mv}:
\begin{equation}
  \begin{array}{ccc}
     \mbox{Primary superconformal fields} & \mbox{weights of }{\cal C} &\mbox{relations}  \\ \hline
    {\cal C}_{\alpha _1\ldots \alpha _m\dot \alpha _1\ldots \dot \alpha _{n}} & (w,c,J_1,J_2) & m=2J_1,\,n=2J_2 \\
     \mbox{real }C_{\alpha _1\ldots \alpha _m\dot \alpha _1\ldots \dot \alpha _{n}}  & (w,0,J,J) &    C={\cal C}={\cal C}^*,\,m=2J  \\
 {\cal D}^\alpha {\cal C}_{\alpha \beta _1\ldots \beta _m\dot \alpha _1\ldots\dot \alpha _n}&(w,c,\ft12(w+c)-1,J_2) &
 m=2J_1-1, 
 \,n=2J_2\\
     {\cal D}_{(\alpha_1 }{\cal C}_{\alpha _2\ldots \alpha _{m)}\dot\alpha _1\ldots\alpha _{n} }&
   (w,c,-\ft12(w+c),J_2) &
    m=2J_1+1,\,n=2J_2\\
    T({\cal C}_{\alpha _1\ldots \alpha _m}) & (w,w-2,J,0) & m=2J
\end{array}
 \label{conditionswcJ}
\end{equation}
The weights of the operations are
\begin{equation}
  {\cal D}_\alpha \ : \ (\ft12,-\ft32,\ft12,0)\,,\qquad\overline{{\cal D}}_{\dot \alpha} \ : \ (\ft12,\ft32,0,\ft12)\,,\qquad
  T\ :\ (1,3,0,0)\,,\qquad \bar T\ :\ (1,-3,0,0)\,.
 \label{weightsoperations}
\end{equation}
Therefore the final weights of some of the expressions in (\ref{conditionswcJ}) are
\begin{equation}
  \begin{array}{cc}
  {\cal D}^\alpha {\cal C}_{\alpha \beta _1\ldots \beta _m\dot \alpha _1\ldots\dot \alpha _n}& (w+\ft12,c-\ft32,\ft12(w+c-3) ,J_2)\,, \\
    {\cal D}_{(\alpha_1 }{\cal C}_{\alpha _2\ldots \alpha _{m)}\dot\alpha _1\ldots\alpha _{n} }& (w+\ft12,c-\ft32,-\ft12(w+c-1) ,J_2)\,, \\
   T({\cal C} _{\alpha _1\ldots \alpha _m})   &  (w+1,w+1,J,0) \,.
   \end{array}
 \label{resultingweights}
\end{equation}

Note that the conditions for the operations ${\cal D}^\alpha {\cal C}_{\alpha \beta _1\ldots \beta _m\dot \alpha _1\ldots\dot \alpha _n}$ and ${\cal D}_{(\alpha_1 }{\cal C}_{\alpha _2\ldots \alpha _{m)}\dot\alpha _1\ldots\alpha _{n} }$
are never compatible. This means that ${\cal D}_\alpha {\cal C}$ only exist if $J_1=0$ in which case the first condition is not applicable. Thus
\begin{equation}
  {\cal D}_\alpha {\cal C}_{\dot \alpha _1\ldots \dot \alpha _m} \mbox{ exists for } V (w,-w,0, J_2) \mbox{ and has }(w+\ft12,-w-\ft32,\ft12,J_2)\,,\qquad m=2J_2\,.
 \label{Dalphaexists}
\end{equation}
If we further want to apply $T$ on ${\cal D}_\alpha {\cal C}$, we need $J_2=0$ and $w=0$. Thus
\begin{equation}
  T({\cal D}_\alpha {\cal C}) \mbox{ exists for } {\cal C} (0,0,0, 0) \mbox{ and has }(\ft32,\ft32,\ft12,0)\,.
 \label{TDalphaexists}
\end{equation}

These restrictions define also which multiplets exist. The first example of this is the (anti)chiral multiplet, which is defined by the vanishing of (\ref{Dalphaexists}), which determines the conditions when this exist. Similarly, a linear multiplet is defined from the condition that $T({\cal L})=0$. In order that this is well defined, ${\cal L}$ should thus have weights $(w,w-2,J,0)$. A real linear multiplet is thus only possible for $(2,0,0,0)$. This gives the following table of multiplets defined by restrictions on a general ${\cal C}$
\begin{equation}
  \begin{array}{cc}
     \mbox{Multiplet} & \mbox{weights} \\ \hline
     \mbox{real} & (w,0,J,J) \\
     \mbox{chiral } & (w,w,J,0)    \\
     \mbox{antichiral} & (w,-w,0,J)    \\
    \mbox{linear} & (w,w-2,J,0) \\
    \mbox{real linear} &(2,0,0,0)\,.
   \end{array}
 \label{multipletweights}
\end{equation}
Densities in the Lagrangian should have weights $(4,0,0,0) $. Therefore the formulas for densities apply for multiplets with weights
\begin{equation}
  [X]_F \ : \ X\mbox{ has } (3,3,0,0),\,\qquad [C]_D \ : \ C\mbox{ has }(2,0,0,0)\,.
 \label{densitiesweights}
\end{equation}

Some further examples of these rules are the following:
\begin{enumerate}
  \item The $T$ operation cannot be applied on a chiral multiplet. On an antichiral multiplet it requires that the latter has $(1,-1,0,0)$.
  \item For a supercurrent we demand that ${\cal D}^\alpha V_{\alpha \beta _1\ldots }$ exists and that $V$ is real. Then $V$ should have $(w,0, \ft12 w -1,\ft12 w -1)$. One can e.g. have $J_1=J_2=\ft12$ and $w=3$.
  \item In order that ${\cal D}^\alpha \psi _\alpha $ exists for a chiral $\psi _\alpha $ we need that $\psi _\alpha $ has weights $(\ft32,\ft32,\ft12,0)$, such that ${\cal D}^\alpha \psi _\alpha $ has $(2,0,0,0)$, which are the weights of a real linear multiplet.
\end{enumerate}

\mciteSetMidEndSepPunct{;\space}{}{\relax}
\bibliography{supergravity}

\providecommand{\href}[2]{#2}\begingroup\raggedright\begin{mcitethebibliography}{10}
\bibitem{Casalbuoni:1988sx}
R.~Casalbuoni, S.~De~Curtis, D.~Dominici, F.~Feruglio  and R.~Gatto,
  \emph{{When does supergravity become strong?}}, Phys. Lett. {\bf B216} (1989)
  \href{http://dx.doi.org/10.1016/0370-2693(89)91123-4}{325},
[Erratum: Phys. Lett.B229,439(1989)]
\bibitem{Kallosh:2000ve}
R.~Kallosh, L.~Kofman, A.~D. Linde  and A.~Van~Proeyen, \emph{Superconformal
  symmetry, supergravity and cosmology}, Class. Quant. Grav. {\bf 17} (2000)
  4269--4338, \href{http://arXiv.org/abs/hep-th/0006179}{{\tt hep-th/0006179}},
erratum \textbf{21} (2004) 5017
\bibitem{Ferrara:2015tyn}
S.~Ferrara, R.~Kallosh  and J.~Thaler, \emph{{Cosmology with orthogonal
  nilpotent superfields}}, Phys. Rev. {\bf D93} (2016), no.~4,
  \href{http://dx.doi.org/10.1103/PhysRevD.93.043516}{043516},
\href{http://arxiv.org/abs/1512.00545}{{\tt arXiv:1512.00545 [hep-th]}}
\bibitem{Carrasco:2015iij}
J.~J.~M. Carrasco, R.~Kallosh  and A.~Linde, \emph{{Inflatino-less Cosmology}},
\href{http://arxiv.org/abs/1512.00546}{{\tt arXiv:1512.00546 [hep-th]}}
\bibitem{Dall'Agata:2014oka}
G.~Dall'Agata and F.~Zwirner, \emph{{On sgoldstino-less supergravity models of
  inflation}}, JHEP {\bf 12} (2014)
  \href{http://dx.doi.org/10.1007/JHEP12(2014)172}{172},
\href{http://arxiv.org/abs/1411.2605}{{\tt arXiv:1411.2605 [hep-th]}}
\bibitem{Volkov:1972jx}
D.~V. Volkov and V.~P. Akulov, \emph{{Possible universal neutrino
  interaction}}, JETP Lett. {\bf 16} (1972) 438--440,
[Pisma Zh. Eksp. Teor. Fiz.16,621(1972)]
\bibitem{Volkov:1973ix}
D.~Volkov and V.~Akulov, \emph{{Is the neutrino a Goldstone particle?}}, Phys.
  Lett. {\bf 46B} (1973)
\href{http://dx.doi.org/10.1016/0370-2693(73)90490-5}{109--110}
\bibitem{Rocek:1978nb}
M.~Ro\v{c}ek, \emph{{Linearizing the Volkov-Akulov model}}, Phys.Rev.Lett. {\bf
  41} (1978)
\href{http://dx.doi.org/10.1103/PhysRevLett.41.451}{451--453}
\bibitem{Ivanov:1978mx}
E.~A. Ivanov and A.~A. Kapustnikov, \emph{{General relationship between linear
  and nonlinear realizations of supersymmetry}}, J. Phys. {\bf A11} (1978)
\href{http://dx.doi.org/10.1088/0305-4470/11/12/005}{2375--2384}
\bibitem{Lindstrom:1979kq}
U.~Lindstr{\"o}m and M.~Ro\v{c}ek, \emph{{Constrained local superfields}},
  Phys.Rev. {\bf D19} (1979)
\href{http://dx.doi.org/10.1103/PhysRevD.19.2300}{2300--2303}
\bibitem{Samuel:1982uh}
S.~Samuel and J.~Wess, \emph{{A superfield formulation of the nonlinear
  realization of supersymmetry and its coupling to supergravity}}, Nucl.Phys.
  {\bf B221} (1983)
\href{http://dx.doi.org/10.1016/0550-3213(83)90622-3}{153}
\bibitem{Casalbuoni:1988xh}
R.~Casalbuoni, S.~De~Curtis, D.~Dominici, F.~Feruglio  and R.~Gatto,
  \emph{{Non-linear realization of supersymmetry algebra from supersymmetric
  constraint}}, Phys.Lett. {\bf B220} (1989)
\href{http://dx.doi.org/10.1016/0370-2693(89)90788-0}{569}
\bibitem{Komargodski:2009rz}
Z.~Komargodski and N.~Seiberg, \emph{{From linear SUSY to constrained
  superfields}}, JHEP {\bf 0909} (2009)
  \href{http://dx.doi.org/10.1088/1126-6708/2009/09/066}{066},
\href{http://arxiv.org/abs/0907.2441}{{\tt arXiv:0907.2441 [hep-th]}}
\bibitem{Kuzenko:2010ef}
S.~M. Kuzenko and S.~J. Tyler, \emph{{Relating the Komargodski-Seiberg and
  Akulov-Volkov actions: Exact nonlinear field redefinition}}, Phys.Lett. {\bf
  B698} (2011)
  \href{http://dx.doi.org/10.1016/j.physletb.2011.03.020}{319--322},
\href{http://arxiv.org/abs/1009.3298}{{\tt arXiv:1009.3298 [hep-th]}}
\bibitem{Antoniadis:2014oya}
I.~Antoniadis, E.~Dudas, S.~Ferrara  and A.~Sagnotti, \emph{{The
  Volkov-Akulov-Starobinsky supergravity}}, Phys.Lett. {\bf B733} (2014)
  \href{http://dx.doi.org/10.1016/j.physletb.2014.04.015}{32--35},
\href{http://arxiv.org/abs/1403.3269}{{\tt arXiv:1403.3269 [hep-th]}}
\bibitem{Ferrara:2014kva}
S.~Ferrara, R.~Kallosh  and A.~Linde, \emph{{Cosmology with nilpotent
  superfields}}, JHEP {\bf 1410} (2014)
  \href{http://dx.doi.org/10.1007/JHEP10(2014)143}{143},
\href{http://arxiv.org/abs/1408.4096}{{\tt arXiv:1408.4096 [hep-th]}}
\bibitem{Kuzenko:2011tj}
S.~M. Kuzenko and S.~J. Tyler, \emph{{On the Goldstino actions and their
  symmetries}}, JHEP {\bf 1105} (2011)
  \href{http://dx.doi.org/10.1007/JHEP05(2011)055}{055},
\href{http://arxiv.org/abs/1102.3043}{{\tt arXiv:1102.3043 [hep-th]}}
\bibitem{Dudas:2015eha}
E.~Dudas, S.~Ferrara, A.~Kehagias  and A.~Sagnotti, \emph{{Properties of
  nilpotent supergravity}}, JHEP {\bf 09} (2015)
  \href{http://dx.doi.org/10.1007/JHEP09(2015)217}{217},
\href{http://arxiv.org/abs/1507.07842}{{\tt arXiv:1507.07842 [hep-th]}}
\bibitem{Bergshoeff:2015tra}
E.~A. Bergshoeff, D.~Z. Freedman, R.~Kallosh  and A.~Van~Proeyen, \emph{{Pure
  de Sitter supergravity}}, Phys. Rev. {\bf D92} (2015), no.~8,
  \href{http://dx.doi.org/10.1103/PhysRevD.92.085040}{085040},
\href{http://arxiv.org/abs/1507.08264}{{\tt arXiv:1507.08264 [hep-th]}}
\bibitem{Hasegawa:2015bza}
F.~Hasegawa and Y.~Yamada, \emph{{Component action of nilpotent multiplet
  coupled to matter in 4 dimensional $ \mathcal{N}=1 $ supergravity}}, JHEP
  {\bf 10} (2015) \href{http://dx.doi.org/10.1007/JHEP10(2015)106}{106},
\href{http://arxiv.org/abs/1507.08619}{{\tt arXiv:1507.08619 [hep-th]}}
\bibitem{Kallosh:2015tea}
R.~Kallosh and T.~Wrase, \emph{{De Sitter supergravity model building}}, Phys.
  Rev. {\bf D92} (2015), no.~10,
  \href{http://dx.doi.org/10.1103/PhysRevD.92.105010}{105010},
\href{http://arxiv.org/abs/1509.02137}{{\tt arXiv:1509.02137 [hep-th]}}
\bibitem{Schillo:2015ssx}
M.~Schillo, E.~van~der Woerd  and T.~Wrase.
\newblock \href{http://arxiv.org/abs/1511.01542}{{\tt arXiv:1511.01542
  [hep-th]}}.
\newblock
To be published by Fortschritte der Physik, 2016.
\bibitem{Kallosh:2015pho}
R.~Kallosh, A.~Karlsson  and D.~Murli, \emph{{From linear to nonlinear
  supersymmetry via functional integration}}, Phys. Rev. {\bf D93} (2016),
  no.~2, \href{http://dx.doi.org/10.1103/PhysRevD.93.025012}{025012},
\href{http://arxiv.org/abs/1511.07547}{{\tt arXiv:1511.07547 [hep-th]}}
\bibitem{Ghilencea:2015aph}
D.~M. Ghilencea, \emph{{Comments on the nilpotent constraint of the goldstino
  superfield}},
\href{http://arxiv.org/abs/1512.07484}{{\tt arXiv:1512.07484 [hep-th]}}
\bibitem{Kahn:2015mla}
Y.~Kahn, D.~A. Roberts  and J.~Thaler, \emph{{The goldstone and goldstino of
  supersymmetric inflation}}, JHEP {\bf 10} (2015)
  \href{http://dx.doi.org/10.1007/JHEP10(2015)001}{001},
\href{http://arxiv.org/abs/1504.05958}{{\tt arXiv:1504.05958 [hep-th]}}
\bibitem{Dall'Agata:2015lek}
G.~Dall'Agata and F.~Farakos, \emph{{Constrained superfields in supergravity}},
  JHEP {\bf 02} (2016) \href{http://dx.doi.org/10.1007/JHEP02(2016)101}{101},
\href{http://arxiv.org/abs/1512.02158}{{\tt arXiv:1512.02158 [hep-th]}}
\bibitem{Kugo:1982cu}
T.~Kugo and S.~Uehara, \emph{{Conformal and Poincar{\'e} tensor calculi in
  $N=1$ supergravity}}, Nucl. Phys. {\bf B226} (1983)
\href{http://dx.doi.org/10.1016/0550-3213(83)90463-7}{49}
\bibitem{VanProeyen:1983wk}
A.~Van~Proeyen, \emph{Superconformal tensor calculus in $N=1$ and $N=2$
  supergravity},
in {\it Supersymmetry and supergravity 1983}, XIXth Winter School and Workshop
  of Theoretical Physics, Karpacz, Poland, ed. B.~Milewski. World Scientific,
  1983
\bibitem{Kugo:1983mv}
T.~Kugo and S.~Uehara, \emph{{$N=1$ Superconformal tensor calculus: multiplets
  with external Lorentz indices and spinor derivative operators}},
  Prog.Theor.Phys. {\bf 73} (1985)
\href{http://dx.doi.org/10.1143/PTP.73.235}{235}
\bibitem{Freedman:2012zz}
D.~Z. Freedman and A.~Van~Proeyen, {\em Supergravity}.
\newblock Cambridge University Press,
2012
\bibitem{Kallosh:2016hcm}
R.~Kallosh, A.~Karlsson, B.~Mosk  and D.~Murli, \emph{{Orthogonal Nilpotent
  Superfields from Linear Models}},
\href{http://arxiv.org/abs/1603.02661}{{\tt arXiv:1603.02661 [hep-th]}}
\bibitem{Dall'Agata:2015zla}
G.~Dall'Agata, S.~Ferrara  and F.~Zwirner, \emph{{Minimal scalar-less
  matter-coupled supergravity}}, Phys. Lett. {\bf B752} (2016)
  \href{http://dx.doi.org/10.1016/j.physletb.2015.11.066}{263--266},
\href{http://arxiv.org/abs/1509.06345}{{\tt arXiv:1509.06345 [hep-th]}}
\bibitem{Ferrara:1999ed}
S.~Ferrara and A.~Zaffaroni, \emph{{Superconformal field theories, multiplet
  shortening, and the AdS$_5$ / SCFT$_4$ correspondence}},
  \href{http://arxiv.org/abs/hep-th/9908163}{{\tt arXiv:hep-th/9908163
  [hep-th]}},
proc. of the Conference Moshe Flato, 5-8 Sep 1999. Dijon, France, {\it
  Quantization, deformations and symmetries}, Ed. G. Dito, D. Sternheimer.
  Kluwer 2000
\bibitem{Ferrara:1978jt}
S.~Ferrara and P.~van Nieuwenhuizen, \emph{Tensor calculus for supergravity},
  Phys. Lett. {\bf B76} (1978)
404
\bibitem{Ferrara:1978wj}
S.~Ferrara and P.~Van~Nieuwenhuizen, \emph{{Structure of Supergravity}}, Phys.
  Lett. {\bf B78} (1978)
\href{http://dx.doi.org/10.1016/0370-2693(78)90642-1}{573}
\bibitem{Stelle:1978wj}
K.~S. Stelle and P.~C. West, \emph{{Relation between vector and scalar
  multiplets and gauge invariance in supergravity}}, Nucl. Phys. {\bf B145}
  (1978)
\href{http://dx.doi.org/10.1016/0550-3213(78)90420-0}{175}
\bibitem{Cremmer:1978iv}
E.~Cremmer, B.~Julia, J.~Scherk, P.~van Nieuwenhuizen, S.~Ferrara  and
  L.~Girardello, \emph{{Super-higgs effect in supergravity with general scalar
  interactions}}, Phys. Lett. {\bf B79} (1978)
\href{http://dx.doi.org/10.1016/0370-2693(78)90230-7}{231}
\bibitem{Cremmer:1978hn}
E.~Cremmer, B.~Julia, J.~Scherk, S.~Ferrara, L.~Girardello  and P.~van
  Nieuwenhuizen, \emph{{Spontaneous symmetry breaking and Higgs effect in
  supergravity without cosmological constant}}, Nucl. Phys. {\bf B147} (1979)
\href{http://dx.doi.org/10.1016/0550-3213(79)90417-6}{105}
\bibitem{vanNieuwenhuizen:1978st}
P.~van Nieuwenhuizen, \emph{{Lectures in supergravity theory}}, in {\em {Recent
  developments in gravitation, Proc. Carg{\`e}se 1978, eds. M.Levy and
  S.Deser}}, pp.~519--548.
\newblock 1979.
\newblock
NATO Sci. Ser. B, vol. 44.
\bibitem{Sohnius:1982xs}
M.~F. Sohnius and P.~C. West, \emph{{Supergravity with one auxiliary spinor}},
  Nucl.Phys. {\bf B216} (1983)
\href{http://dx.doi.org/10.1016/0550-3213(83)90489-3}{100}
\bibitem{Cremmer:1982wb}
E.~Cremmer, S.~Ferrara, L.~Girardello  and A.~Van~Proeyen, \emph{{Coupling
  supersymmetric Yang--Mills theories to supergravity}}, Phys. Lett. {\bf B116}
  (1982)
\href{http://dx.doi.org/10.1016/0370-2693(82)90332-X}{231}
\bibitem{Cremmer:1982en}
E.~Cremmer, S.~Ferrara, L.~Girardello  and A.~Van~Proeyen, \emph{{Yang--Mills
  theories with local supersymmetry: Lagrangian, transformation laws and
  superhiggs effect}}, Nucl. Phys. {\bf B212} (1983)
\href{http://dx.doi.org/10.1016/0550-3213(83)90679-X}{413}
\bibitem{VanProeyen:1979ks}
A.~Van~Proeyen, \emph{{Massive vector multiplets in supergravity}}, Nucl. Phys.
  {\bf B162} (1980)
\href{http://dx.doi.org/10.1016/0550-3213(80)90345-4}{376}
\bibitem{Cecotti:1987sa}
S.~Cecotti, \emph{{Higher derivative supergravity is equivalent to standard
  supergravity coupled to matter. 1.}}, Phys.Lett. {\bf B190} (1987)
\href{http://dx.doi.org/10.1016/0370-2693(87)90844-6}{86}
\bibitem{Bergshoeff:2016psz}
E.~Bergshoeff, D.~Freedman, R.~Kallosh  and A.~Van~Proeyen, \emph{{Construction
  of the de Sitter supergravity}},
\newblock 2016.
\newblock \href{http://arxiv.org/abs/1602.01678}{{\tt arXiv:1602.01678
  [hep-th]}}.
\newblock
  \url{http://inspirehep.net/record/1419685/files/arXiv:1602.01678.pdf}.
\newblock
To be published in the proceedings of the workshop `About various kinds of
  interactions' in honor of Philippe Spindel, Mons, June 2015.
\bibitem{deWit:1981fh}
B.~de~Wit and M.~Ro\v{c}ek, \emph{{Improved tensor multiplets}}, Phys.Lett.
  {\bf B109} (1982)
\href{http://dx.doi.org/10.1016/0370-2693(82)91109-1}{439}
\bibitem{Ferrara:1983dh}
S.~Ferrara, L.~Girardello, T.~Kugo  and A.~Van~Proeyen, \emph{Relation between
  different auxiliary field formulations of $N=1$ supergravity coupled to
  matter}, Nucl. Phys. {\bf B223} (1983)
191
\bibitem{Kugo:2016zzf}
T.~Kugo, R.~Yokokura  and K.~Yoshioka, \emph{{Component versus superspace
  approaches to $D=4$, ${\cal N}=1$ conformal supergravity}},
\href{http://arxiv.org/abs/1602.04441}{{\tt arXiv:1602.04441 [hep-th]}}
\end{mcitethebibliography}\endgroup
\bibliographystyle{toinemcite}

\end{document}